\documentclass[]{tPHM2e_mod}

\begin{document}
\DeclareGraphicsExtensions{.jpg,.pdf}
%\doi{10.1080/1478643YYxxxxxxxx}
%\issn{1478-6443}
%\issnp{1478-6435}
\jvol{00} \jnum{00} \jyear{2009} \jmonth{31 May}

\markboth{R. E. Rudd}{Void Growth in BCC Metals}

\articletype{ARTICLE}

\title{Void Growth in BCC Metals Simulated with Molecular Dynamics using the Finnis-Sinclair Potential}

\author{Robert E. Rudd$^{\ast}$
\thanks{$^\ast$Corresponding author. Email: robert.rudd@llnl.gov}
\\\vspace{6pt}  {\em{Lawrence Livermore National Laboratory, Livermore, CA 94550-9698 USA}}\\\vspace{6pt}\received{31 May 2009} }

\maketitle

\begin{abstract}
The process of fracture in ductile metals involves the nucleation, growth,
and linking of voids.  This process takes place both at the low rates 
involved in typical engineering applications and at the high rates 
associated with dynamic fracture processes such as spallation.  Here
we study the growth of a void in a single crystal at high rates 
using molecular dynamics (MD) 
based on Finnis-Sinclair interatomic potentials for the
body-centred cubic (bcc) metals V, Nb, Mo, Ta, and W.  
The use of the
Finnis-Sinclair potential enables the study of plasticity associated
with void growth at the atomic level at room temperature and strain
rates from 10$^9$/s down to 10$^6$/s and systems as large as
128 million atoms.  The atomistic systems are 
observed to undergo a transition from twinning at the higher end
of this range to dislocation flow at the lower end.  We analyze
the simulations for the specific mechanisms of plasticity 
associated with void growth as dislocation loops are punched out
to accommodate the growing void.  We also analyse the process
of nucleation and growth of voids in simulations of nanocrystalline 
Ta expanding at different strain rates. 
We comment on differences in the plasticity associated with void
growth in the bcc metals compared to earlier studies in face-centred
cubic (fcc) metals. 
%(must be $<$ 250 words)
\bigskip

\bigskip

\end{abstract}

\section{Introduction}

The process of ductile failure in metals involves the nucleation and 
growth of voids \cite{Curran}.  
Formation of the voids locally relieves the elastic 
strain energy built up in the material under tension.  The voids link 
up to form the fracture surface, which further relieves the strain energy.  
Voids nucleate at the weak points in the material when the tensile stress 
exceeds the material strength at the nucleation site.  
Often these nucleation sites are at weakly bound inclusions or grain 
boundary junctions.  The material around growing voids undergoes a 
large deformation (strains of order unity) in order to accommodate 
the growing void.  These large strains require plastic deformation of the 
matrix material around the void. The interaction of the porosity 
characteristic of damage and ductile failure and the concomitant 
matrix plasticity has been studied theoretically using continuum 
\cite{McClintock2,RiceTracey,tvergaard84,Giessen,Wu},
mesoscale \cite{Stevens},
and atomistic \cite{RuddJCAMD,belak2,RuddBelak} techniques. 
Mesoscale \cite{Wolfer,MeyersVoid,SofronisVoid} and 
continuum \cite{hill,Gurson} 
models of void growth and damage have been developed,
and yet many aspects remain poorly understood. 

One aspect of continued interest is the effect of high strain rates.
In engineering applications and mechanical 
tests the strain rates involved in fracture are low; however, in processes 
such as projectile penetration and fragmentation ductile fracture occurs 
under dynamic conditions with very high strain rates \cite{McClintock}. 
An interesting
recent example of a dynamic fracture application is the fragmentation
of structures in fusion-class laser systems due to stray light in 
the enormously powerful lasers \cite{Koniges}.  
Controlled dynamic fracture experiments can be conducted on a variety
of platforms such as gas guns, split Hopkinson bars and lasers.  
Typically, a plane-fronted compressive wave is generated.  
It propagates through the sample and reflects off a free surface.
The compressive wave become tensile as it reflects off the surface.
If the tension exceeds the material strength, fracture occurs.  This
kind of dynamic fracture is known as spallation.

In this article we study void growth in ductile body-centred
cubic (bcc) metals that is associated with dynamic fracture
such as spallation.  We do not model the wave propagation
explicitly \cite{Bringa,GermannSpall}; 
rather we use molecular dynamics (MD) simulation
to capture the salient features of the process in a representative
volume element that is small compared to the thickness of the
rarefaction wave at the spall plane.  There are merits to
a direct simulation of the wave propagation, but it is challenging
in MD to have a simulation sufficiently large in spatial extent
and that runs sufficiently long to establish a self-similar
wave.  Also, the wave form is affected by plastic flow that is
initially independent of the growth of small voids since
the yield stress is typically significantly below the
threshold for nucleation of dislocations at the void
surfaces.  It is challenging to capture realistic length
scales of the dislocation density in MD.  So we focus on the
case where the release waveform is specified, with a wave front
large compared to the voids.  We also assume that plastic
flow has eliminated much of the shear component of the wave,
and we therefore treat the ideal case of equiaxed (hydrostatic)
expansion.  We consider two cases: one where the voids nucleate 
from weakly bound inclusions and another where they nucleate
from grain boundary junctions in a nanocrystalline system.
At low rates a few low-threshold sites can nucleate voids 
that relieve the tensile stress in a relatively large volume
around them; at higher rates, the tensile stress rises to the
point that voids nucleate from higher threshold sites before
the relief wave from the weakest sites reach them.  The result
is a higher density of nucleation sites with smaller, stronger
nuclei.  

Our work here builds on our previous studies of void processes
in the dynamic fracture of face-centred cubic (fcc) metals \cite{RuddJCAMD}.
Using molecular dynamics simulations of fcc copper systems, we
have studied the threshold for void nucleation and the subsequent
growth and linking processes in nanocrystalline copper \cite{RuddBelak}.
Prismatic dislocation loops punched out to accommodate void growth in 
an MD simulation were first shown in Ref.\ \cite{Moriarty2002}.
We have shown how varying the stress triaxiality from uniaxial
to triaxial tension affects the void morphology and
plasticity during growth \cite{SeppalaTriaxPRB,disl,DupuySurf}.
We have also examined how voids interact with each other
as they transition from a regime of independent growth
through the beginning of the coalescence process in which
neighbouring voids expand more rapidly toward each other
\cite{SeppalaCoalescePRL,SeppalaCoalescePRB}.

Here we extend the investigations to bcc metals for several reasons.  
The higher Peierls barrier for dislocation motion in bcc metals leads 
to some interesting differences in the mechanisms active in the plastic 
zone surrounding a void.  Also, the bcc dislocations are not split
into partials, so the dislocation dynamics is not affected by
stacking faults as it is in the fcc case.  The behaviour of 
nanocrystalline bcc metals, including their ductility and
fracture, is of current interest \cite{vanadiumMa,Hamza}. The generation of
copious debris has been reported in MD simulations of dislocation
motion in bcc metals \cite{MarianBulatov}, and it is interesting
to see whether such a process plays are role in bcc void growth.
Furthermore, bcc metals are known to be prone to twinning at
high rate deformation \cite{Christian}, and the role of twinning 
in void growth has not been explored.
A considerable quantity of experimental data on spallation 
in the bcc metal tantalum is available \cite{Gray}.
Most continuum models of damage do not explicitly account
for the lattice structure.  The lattice structure only matters
insofar as it affects the yield surface \cite{hill,Gurson}.
Marked differences observed in the behaviour of voids in bcc and fcc
metals then points to new directions for constitutive model
development and validation experiments at large laser 
facilities and/or fourth generation light sources.

We are writing this article as part of a commemoration of the twenty-fifth
anniversary of the publication of the Finnis-Sinclair potential \cite{FS}.  
Finnis-Sinclair potentials have proved to be extremely valuable 
for the study of plasticity at the atomistic level.  Based on 
the physics of electronic structure at the level of the second
moment approximation to tight-binding theory \cite{SuttonTBM},
they provide a robust description of the bonding physics.
At the same time they neglect much of the rococo detail of
mixed nearly-free-electron and covalent bonding characteristic
of transitions metals that is captured by higher-moment
contributions and angular forces. As a result of this 
approximation, they are computationally efficient, 
many times less expensive than 
fourth-moment bond order potentials \cite{Pettifor,Vitek1,Vitek2}
and Model Generalised Pseudopotential Theory interatomic
potentials \cite{Moriarty1,Moriarty2}.  This reduction in 
computational overhead facilitates the simulation of reasonably 
large system sizes in a way that captures the 
principal features of the physics relevant to plasticity.
It permits relatively quick exploration of large parameter 
spaces, in some cases suggesting the use of more expensive
potentials to determine specific details.

As an extreme example of the horizons opened by these 
inexpensive and robust potentials, we recently used an
aluminium-copper Finnis-Sinclair potential \cite{MasonRuddSutton}
to simulate flows
of up to 62.5 billion ($6.25\times 10^{10}$) atoms for over a nanosecond
in order to
capture both atomic level phenomena and micron-scale hydrodynamic
effects in a complex fluid flow \cite{Gloslietal}.  
It was also
shown how to implement a `onepass' time integrator that further
sped up Finnis-Sinclair simulations by nearly a factor of two,
achieving an average performance of 104~Tflop/s in a full
simulation.  One continuous
2D simulation required 1800 CPU-years, the largest continuous computer
simulation ever to the best of our knowledge. This simulation was
restarted and ultimately ran for over 2800 CPU-years.  The 3D simulation,
although not reported yet, took considerably more computer resources.
The ability to evolve a cubic micron of atoms for over a nanosecond
of simulated time simply would
not have been possible with more expensive potentials, nor would it
have been worthwhile with less reliable potentials.  
The simulations we report here are more modest in size and
computer requirements with no more than 128 million atoms, but they 
are enabled by the same advantageous
features of the Finnis-Sinclair potentials.

The article is organized as follows.  The simulation methods are
presented in Section \ref{sec-methods}. The single crystal simulations
differ somewhat from the nanocrystalline simulations, and the details
of both series of simulations are described.  In Section \ref{sec-analysis}
the techniques used to analyse the large-scale simulations are presented,
including the centrosymmetry analysis used for the dislocations and
the orientation analysis used for the nanocrystalline deformation.
Section \ref{sec-stressStrain} describes the stress-strain behaviour
of the metals under tension as void growth commences.
Section \ref{sec-disl} describes the process of dislocation nucleation
and glide as the matrix material surrounding the void deforms
plastically to accommodate the expanding void.
In Section \ref{sec-fcc}
the results of the simulations of void growth in bcc metals
are compared to our prior findings for void growth in fcc metals.
Section \ref{sec-polycrystal} describes the nucleation and growth
of voids in nanocrystalline Ta.  Finally, our conclusions are
presented in Section \ref{sec-conclusion}.

\section{Simulation Methods}
\label{sec-methods}

We have conducted the void simulations with the FEMD code, a hybrid
finite element (FEM) and MD code for distributed memory
supercomputers \cite{CLS,CGMD}, running in pure 
molecular dynamics mode. We have used the Finnis-Sinclair potentials
\cite{FS} with the Ackland-Thetford core correction 
\cite{AT} for the bcc transition metals V, Nb, Mo, Ta, and W.
We report simulations of void growth in single crystal
metals as well as void nucleation and growth in nanocrystalline
Ta.  The simulations differed somewhat in the two cases,
as described below.

We have also used FEMD to simulate void growth with the part of the
lattice near the void represented with MD and the periphery
represented with FEM.  This approach is similar to the technique
used by Marian et al.\ \cite{MarianVoid}.  The advantage of the 
hybrid MD/FEM technique is to reduce the computational cost of
the large volume of the system that is just elastic.  We do
not report the results of those simulations here.

\subsection{Single Crystal Void Growth Simulation}

The single crystal simulations begin with the simulation
box filled with a single crystal of the transition metal
with a pre-existing void in the centre.  The simulation
box is expanded at a fixed rate, putting the material
under tension, and the void response is monitored.  The
basic elements of this void simulation technique were introduced
by Belak \cite{belak2}.
Specifically, we use a cubic simulation box with periodic boundary 
conditions with an atomic configuration 
in a single crystal bcc lattice, unless noted otherwise.
A crystal lattice with $60^3$ unit cells was constructed
and then all of the atoms were removed inside a spherical void at
the centre of the simulation box with
radius equal to one tenth of the simulation box size.
After the void was formed, 430~195 atoms remain in the box.
In some cases a simulation with $100^3$ unit cells 
and 1~991~622 atoms was used. 
We have also run a series of simulations with $400^3$ unit cells 
with a total of 127~463~891 atoms.
It will be
noted below when these 2-million-atom and 128-million-atom simulations 
were used.
The lattice was oriented with the crystal $<$100$>$ directions
aligned with the simulation box $<$100$>$ directions.
The simulations were carried out starting with the
atoms thermalised at room temperature (300K) and zero pressure.
The system was thermalised using velocity renormalisation
every 100 time steps on the system with the void, and the
simulation box was scaled to zero pressure.
The actual initial size of the box depends on the zero pressure
lattice constant for the metal: 
$a_V=3.0399$ {\AA}, 
$a_{Nb}=3.3008$ {\AA}, 
$a_{Mo}=3.1472$ {\AA}, 
$a_{Ta}=3.3058$ {\AA}, and
$a_W=3.1652$ {\AA}.

To simulate void growth in release wave conditions, we focus
on a representative volume element that is expanding at a 
high rate.  The simulation box is made to
expand at the specified strain rate
equally in all three dimensions.  We use the technique of
Parrinello and Rahman \cite{Parrinello}, writing the
atomic coordinates ${\mathbf{r}}_i$ in terms of a metric
$h_{\alpha \beta}$ times scaled coordinates ${\mathbf{s}}_i$
that take on values in the unit cube:
\begin{equation}
r_{i\alpha} = \sum _{\beta} h_{\alpha \beta} s_{i\beta}
\end{equation}
where the Greek indices indicate spatial dimensions
and lower case Latin indices indicate atom number.
The metric is increased at a specified true strain
rate.  The use of uniformly scaled coordinates prevents
shock wave formation that would occur if the expansion
were driven by tractions on the exterior of the
simulation box.

The equations of motion ($F=ma$) were integrated using a 
velocity Verlet integrator \cite{AllenTildesley}.  
The time step was 3~fs for
tantalum, and scaled for the other metals according to
the ratio of the lattice constant to the speed of sound.
No thermostat was used during the box expansion.  In
practice the temperature dropped by $\sim$20K during
the elastic phase of the expansion and then increased
by a few hundred degrees during the plastic flow.

\subsection{Nanocrystalline Ta Void Simulation}

In the case of nanocrystalline Ta the basic technique
was the same: a simulation box with periodic boundary
conditions was expanded at a fixed strain rate.  
All three dimensions were expanded equally.
The initial configuration of the atoms was the principal
difference.  A fully dense nanoscale polycrystalline Ta system
consisting of 16~384~000 atoms
was used as the initial configuration.  The nanocrystalline
configuration was produced in the simulations of
Streitz et al.\ \cite{StreitzSolid}, in which molten Ta
was compressed at a high strain rate until it solidified.
These simulations were very intensive computationally \cite{StreitzGB}.
We quenched this system down to low temperature and
expanded the box until the pressure was approximately
zero.  We then equilibrated the system to a temperature of
300K and a pressure of 2.55~GPa using velocity renormalisation and
rescaling the box \cite{Parrinello}.  In prior work we have
also studied the behaviour of this system under shear
deformation \cite{RuddMSF}.

The Finnis-Sinclair potentials are very efficient computationally.
Most of the half-million-atom simulations were sufficiently
inexpensive to be conducted on a single processor.
The larger simulations were run on supercomputers.
A 16-million Ta atom configuration on 16 nodes (128 processors)
of the Atlas supercomputer at LLNL \cite{LC}
took an average of 1.3 seconds per time step 
including all I/O and on-the-fly dislocation and orientation analyses.  
The largest simulation took an average of 0.24 seconds per time step for 
128 million Mo atoms on 512 nodes (4096 processors) of Atlas even though
there was a load imbalance across the processors due to the
voided region.  In all of the simulations we have used a uniform
Cartesian spatial domain decomposition.

\section{Analysis Methods}
\label{sec-analysis}

Molecular dynamics produces a different spatial configuration of the
atoms at each time step. The simulations reported here contain
millions of atoms.  Much of the information in the atomic
coordinates is irrelevant to an analysis of the mechanisms of
void growth and plasticity.  A reduction of these data to a
more manageable form is needed, and it is preferable to effect
that reduction as the code runs, on the fly, so that only a 
subset of the coordinates need be written to disk.  The
reduction saves disk space, code execution time (the write
time can be significant), and analysis time (the time to
read the irrelevant data into an analysis routine is also
significant).

\subsection{Dislocation Identification}

\begin{figure}
\begin{center}
\begin{minipage}{100mm}
\subfigure[]{
\resizebox*{70mm}{!}{\includegraphics{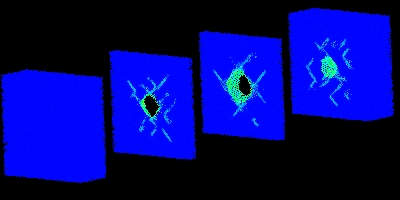}}}%
\\
\subfigure[]{
\resizebox*{70mm}{!}{\includegraphics{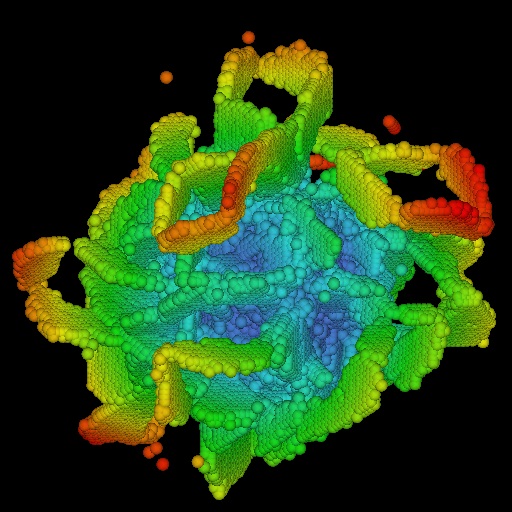}}}%
\caption{(a) Simulation box sliced into four parts 
to show the void explicitly and coloured to show dislocations 
around the void; (b)
visualisation showing only those atoms at defective sites
including dislocations, the void surface, and stacking faults.  
Atoms are coloured according to their distance from the centre of the void.
The parallelogram structures are prismatic loops with stacking
faults on each side \cite{Moriarty2002}.
Both images are from simulations of void growth in copper,
an fcc transition metal.}
\label{fig-vis}
\end{minipage}
\end{center}
\end{figure}

The centrosymmetry deviation has proved to be an effective 
means of finding dislocations.  Introduced by Kelchner
et al.\ \cite{Kelchner} for dislocation analysis in molecular statics
calculations for fcc crystals,
we extended it for molecular dynamics at finite 
temperature \cite{Moriarty2002}. 
The basic algorithm of calculating the mean-square
deviation from centrosymmetry at atom $j$, $\gamma _j$,
\begin{equation}
\gamma _j = \sum _{pairs: i,i'} \left| {\mathbf u}_i + {\mathbf u}_{i'} \right| ^2
\end{equation}
provides a structure-based determination of defects that is
fairly insensitive to temperature, even as originally formulated \cite{Kelchner}
where the displacement is with respect to the
central atom ${\mathbf u}_i={\mathbf x}_i - {\mathbf x}_j$.
The pairs are formed by matching each neighbour atom $i$ 
with the atom $i'$ most nearly centrosymmetric to it
with respect to the atom $j$.  We assign each atom
uniquely to a pair, pairing the nearest neighbour first
and continuing in order of proximity to atom $j$.
Determining defect locations by structure rather than
atomic energies \cite{Zhou,Fabraham} is less sensitive
to thermal noise.
For moderate temperatures ($T<T_{melt}/2$), the $\gamma _j$
is even less sensitive to temperature if the displacements ${\mathbf u}_i$
are taken with respect to the centre of mass of the 
neighbour cluster rather that with respect to atom $j$ \cite{Moriarty2002}. 
Then large values of $\gamma _j$ are indicative of defects
such as dislocation cores and stacking faults,\footnote{Summing
just the three pairs with the 
lowest $\left| {\mathbf u}_i + {\mathbf u}_{i'} \right| ^2$ 
gives a variant of centrosymmetry deviation that detects
the fcc dislocation cores but not the stacking faults.}
and plotting those atoms with $\gamma _j$ greater than a
threshold value provides a means of visualising the dislocation
structure, as depicted in Fig.\ \ref{fig-vis}.

For fcc metals, it works well to sum over the six pairs that
correspond to the 12 nearest neighbours in the perfect fcc
lattice.  For bcc metals, we sum over the seven pairs that
correspond to the first and second neighbour shells in the
perfect bcc lattice.  This choice has proved to be less
noisy than just summing over the four pairs suggested
by the first neighbour shell.  Similarly, in silicon
the centrosymmetry deviation is naturally defined by
the sum over six pairs skipping the four nearest neighbours,
effectively defining $\gamma _j$ on an fcc sublattice
and ignoring the non-centrosymmetric nearest neighbours.
For the bcc visualisations below, we sum seven pairs.
The centrosymmetry deviation is determined from differences
in atomic positions, so it scales with the
nearest neighbour spacing.  It is possible to eliminate this
material dependence with a rescaling of the displacements
by the lattice constant.

\subsection{Orientation Identification}

In the case of the nanocrystalline simulations it is also
useful to have a means of determining the local lattice
orientation.  The orientation provides a way to tell
grains apart.  Furthermore changes in orientation indicate
plastic deformation by dislocation, grain rotation and
twinning.  We have recently introduced a technique to
determine the lattice orientation on the fly during a
simulation, writing it in the form of a quaternion to
enable further analysis, as described in Ref.\ \cite{RuddMSF}.
We direct the reader to that reference for the details of 
the technique, and here limit the discussion to a brief 
comment about quaternions.  
Lattice orientation like other rigid body rotations is specified 
by three numbers such as the Euler angles \cite{Goldstein}. Quaternions provide
another way to describe the orientation. A quaternion may be expressed 
as a 4-vector $(q_0,q_1,q_2,q_3)$ with the constraint that $\sumΣq_i^2 = 1$ 
for unit quaternions.  Quaternions obey multiplication rules appropriate 
for rotations; i.e.\ they form a representation of the rotation group SO(3). 
Unlike the Euler angles they provide a parameterisation of the orientation 
of a rigid body that is free from coordinate singularities 
\cite{Grimmer,DJEvans,AllenTildesley}. 
The visualisation of orientation is facilitated using a colouring 
that maps the three independent parameters, say $(q_1,q_2,q_3)$ to
the intensities of red, green, and blue \cite{RuddMSF}.

\section{Stress-Strain Response}
\label{sec-stressStrain}

One of the first quantities to study in dynamic fracture is
the stress-strain response.  In theory it forms the basis
for constitutive models of void growth and damage.
In experiment the principal features of 
this response can be obtained from surface velocity measurements
in dynamic experiments using techniques like Velocity 
Interferometer System for Any Reflector (VISAR). 
The quantity most often reported to quantify dynamic fracture
is the spall strength, a measure of the tensile stress needed for
spallation determined from the pull-back part of the surface 
velocity (VISAR) curve \cite{MeyersBook}.
The MD simulations provide a complete
stress-strain curve for the sample, and the peak tensile stress
can be compared with experimental spall strength measurements.

We have computed the stress in the MD simulation using the
virial expression:
\begin{equation}
\sigma_{\alpha \beta} = - \frac{1}{V} \left( \sum_i m_i \dot{r}_{i \alpha}
\dot{r}_{i \beta}  + \sum_i \sum_{j>i} r_{ij \alpha} f_{ij \beta} \right).
\label{eq_stresstensor}
\end{equation}
The first term in the stress tensor is the kinetic contribution of
atoms denoted with $i$ and having masses $m_i$ and velocities 
$\dot{{\mathbf{r}}}_i$. The
second term, a microscopic virial potential stress, consists of
sums of interatomic forces ${\mathbf{f}}_{ij}$ of atom pairs 
$\langle ij\rangle$
with corresponding interatomic distances ${\mathbf{r}}_{ij}$.  
The indices $i$ and $j$ denote the atoms, and the indices 
$\alpha$ and $\beta$ denote the Cartesian directions.  The thermal
stress is included, although in practice it makes only a minor
contribution to the changes in stress during the simulated deformation.  
For a recent discussion
of stress calculations in atomistic systems see Ref.\ \cite{Zimmerman}.
The total system volume is given by $V$.  It includes the
volume of the void, so the actual stress in the material is
greater in magnitude than the value reported. For elastically
isotropic continua the stress tensor at point ${\mathbf{r}}$ varies
around a spherical void of radius $R$ subjected to a (negative) pressure $P$ 
according to the form
\begin{equation}
{\mathbf{\sigma}} ({\mathbf{r}}) = -P \, {\mathbf{I}} + P \left(
\frac{R}{|{\mathbf{r}}|} \right)^3  \hat{\mathbf{r}} \otimes \hat{\mathbf{r}}
\label{eq-stress}
\end{equation}
where $\hat{\mathbf{r}}$ is the radial unit vector \cite{Love}.
The stress field for more general cases can be determined
using Eshelby's techniques \cite{Eshelby}.
In any case, the stress varies in the vicinity of the void.
The stress value we report should be considered an effective
stress for the representative volume element containing the
void, averaging over these variations.

\begin{figure}
\begin{center}
\begin{minipage}{140mm}
\subfigure[]{
\resizebox*{70mm}{!}{\includegraphics{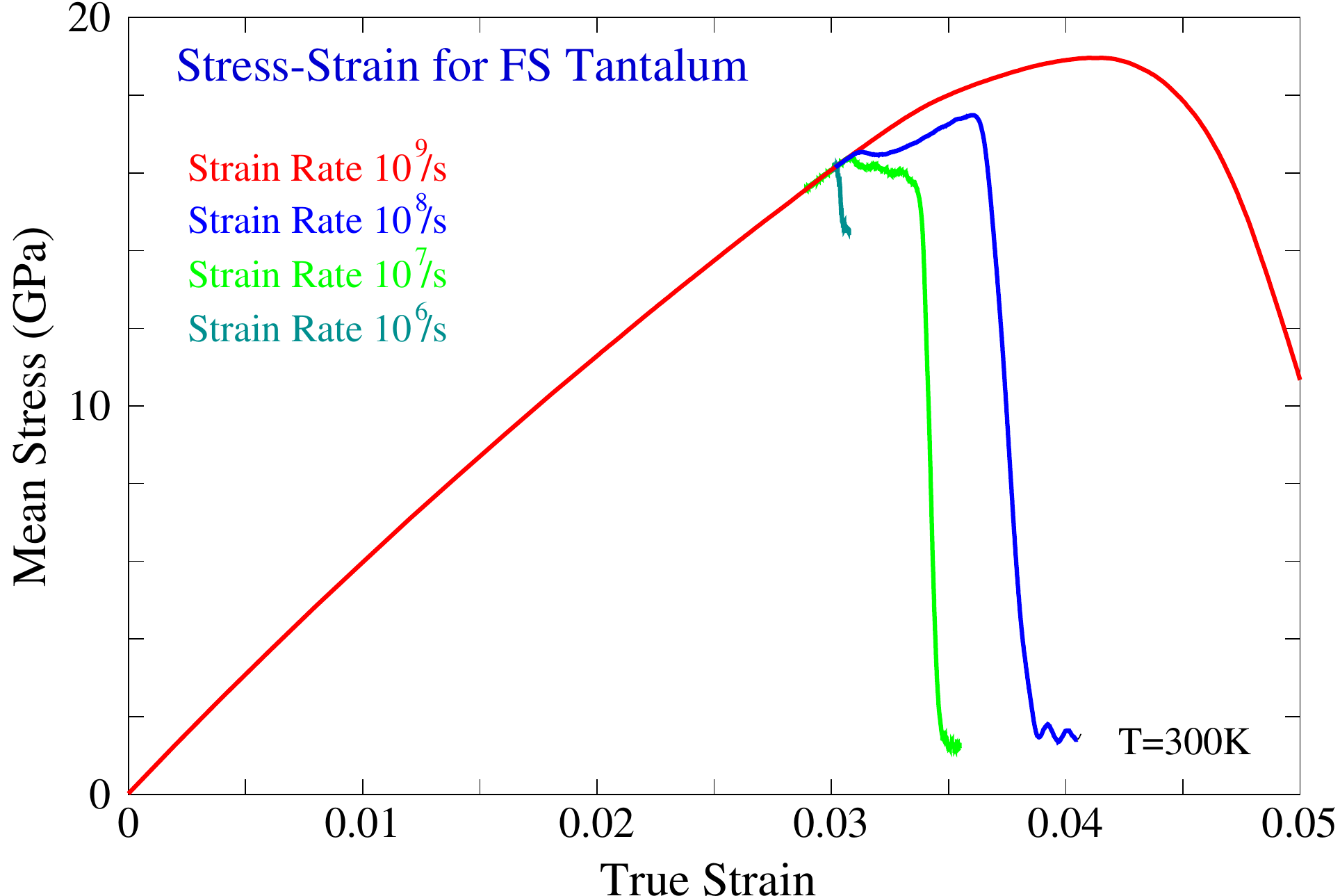}}}%
\subfigure[]{
\resizebox*{70mm}{!}{\includegraphics{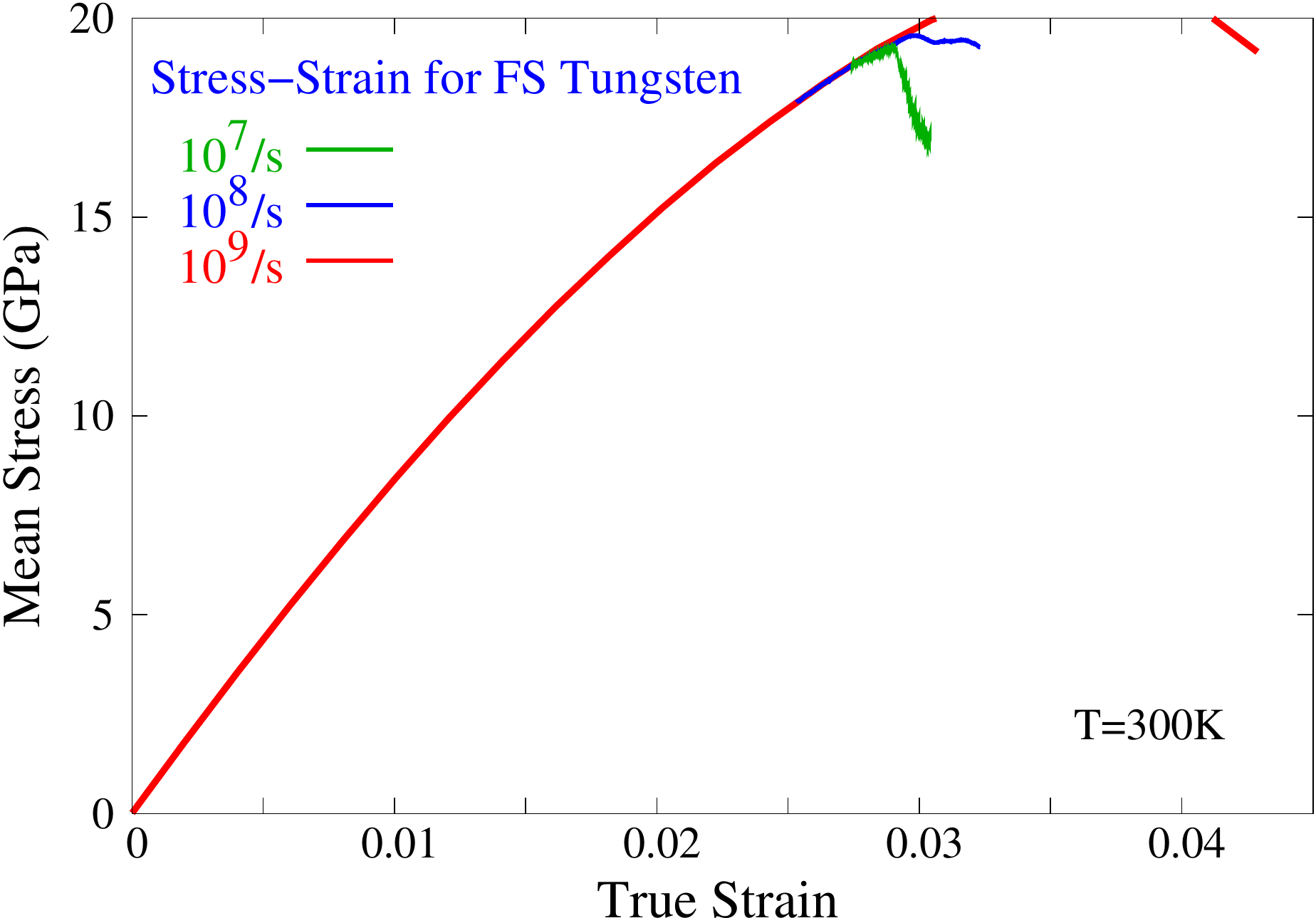}}}%
\\
\subfigure[]{
\resizebox*{66mm}{!}{\includegraphics{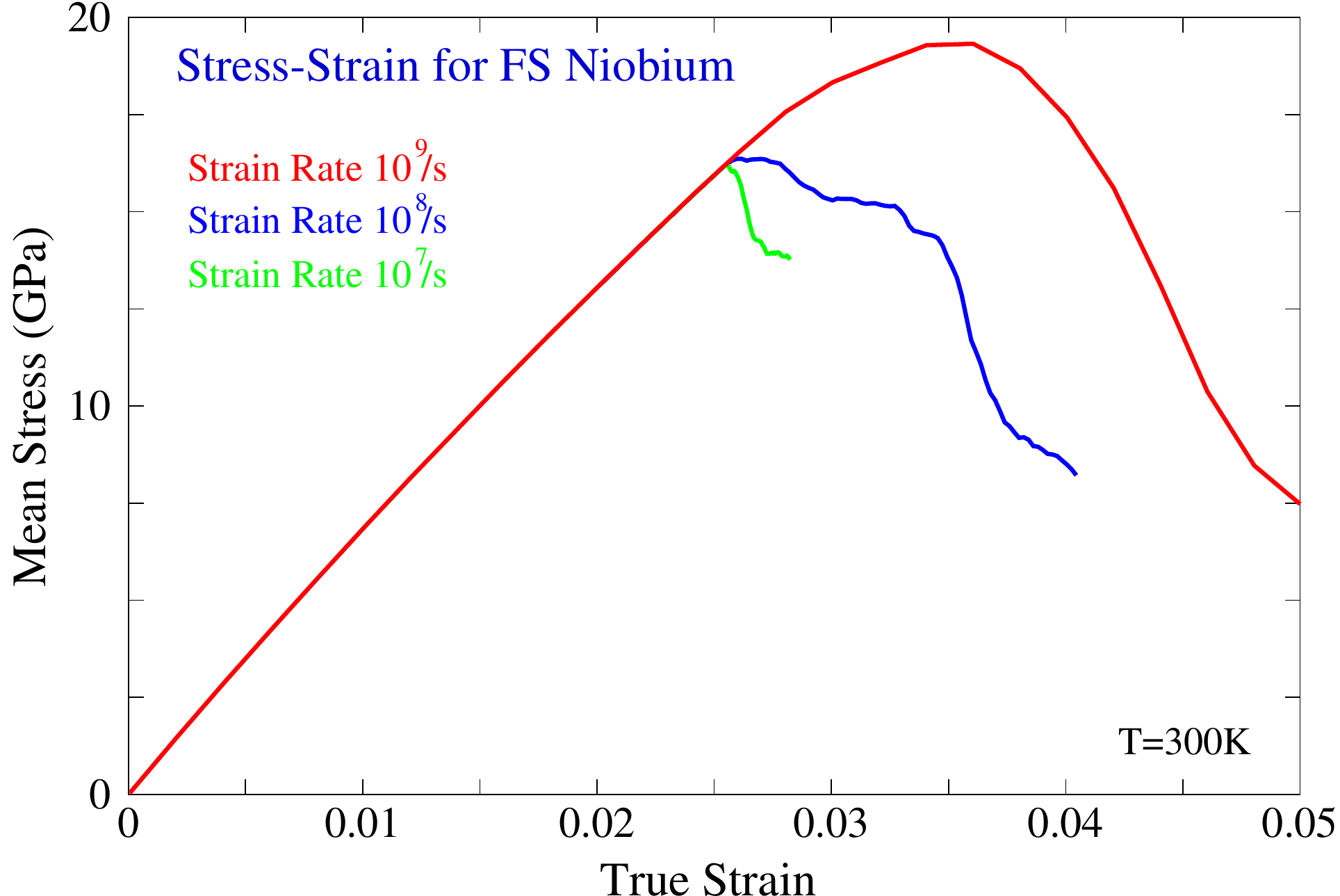}}}%
\subfigure[]{
\resizebox*{74mm}{!}{\includegraphics{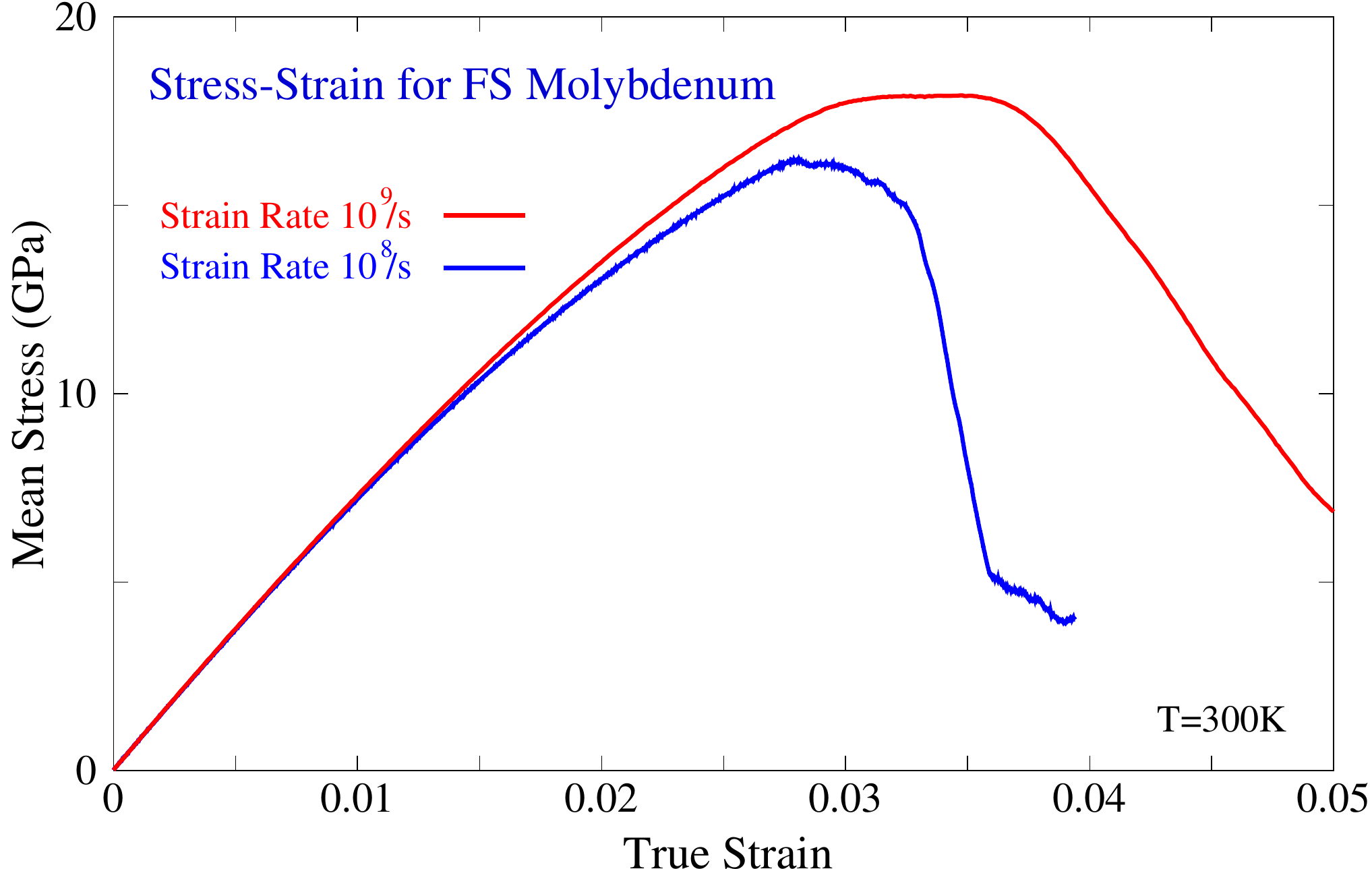}}}%
\\
\subfigure[]{
\resizebox*{70mm}{!}{\includegraphics{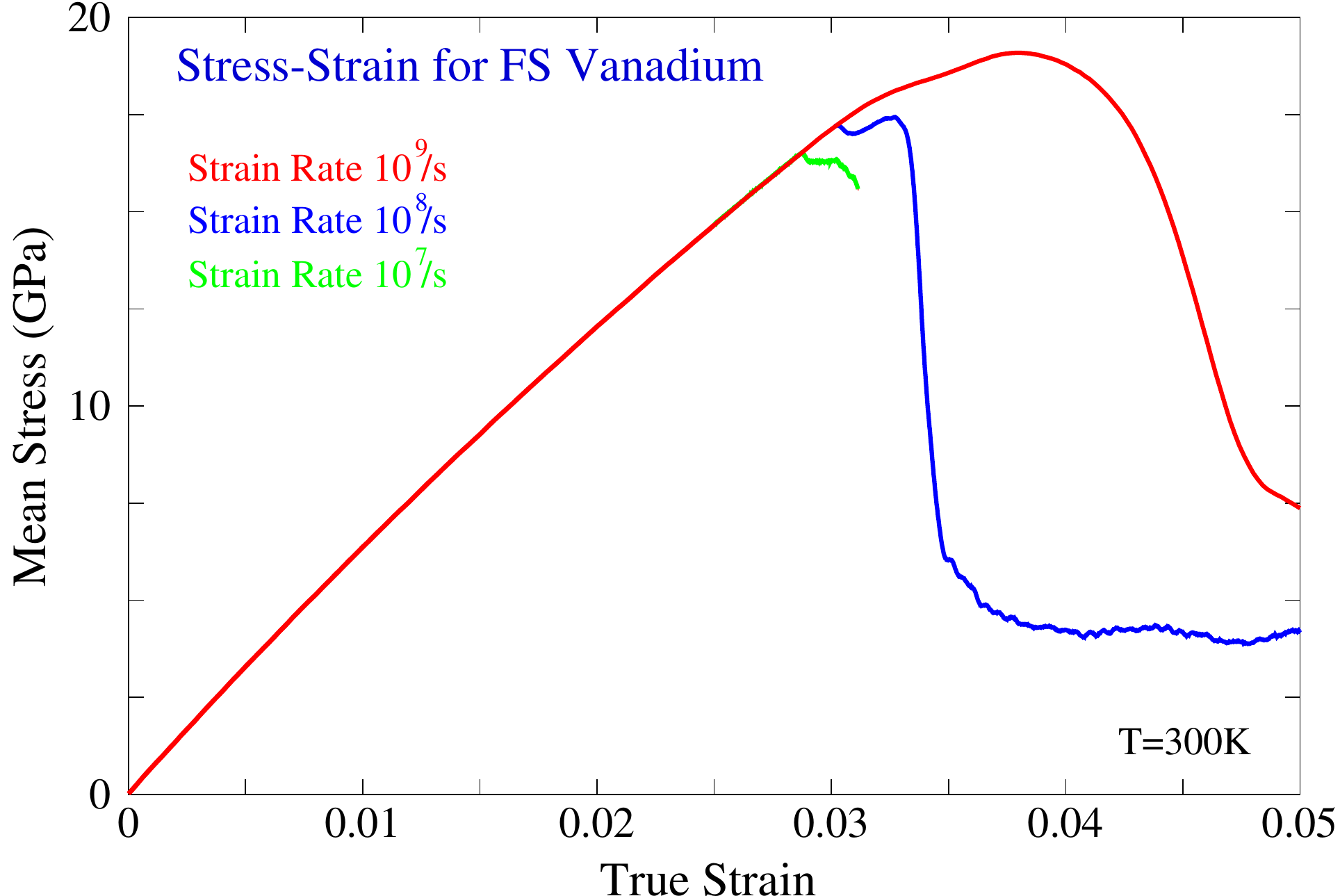}}}%
\caption{Stress-strain responses for void growth in
five bcc metals as simulated in MD using Finnis-Sinclair (FS)
potentials:  (a) tantalum, (b) tungsten (c) niobium,
(d) molybdenum and (e) vanadium.  The 
different curves in each plot correspond to different strain
rates, as indicated. All of the
simulations began at room temperature.  
The mean stress, plotted on the vertical axis, 
is the negative of the pressure.
\label{fig-stress}
}
\end{minipage}
\end{center}
\end{figure}

The stress-strain curves for the group VA bcc transition
metals (V, Nb, Ta) and VIA bcc transition metals (Mo and W) 
are shown in Fig.\ \ref{fig-stress}.  In each case the
simulation box was expanded at a specific true strain
rate, expanding the three dimensions equally.  The
mean stress, $\frac{1}{3} {\mathrm Tr}~\sigma$, is plotted as a
function of the true strain $\log L/L_0$, where
$L$ and $L_0$ are the current and zero-strain box sizes, respectively.
Stress is plotted using the engineering convention that
positive stress corresponds to tension.
The stress-strain curves at lower strain rates are terminated
earlier than at higher strain rates to limit the computational
expense which scales as one over the strain rate.  A simulation
at $\dot{\varepsilon}=10^6$/s is 1000 times as expensive as
a simulation at $\dot{\varepsilon}=10^9$/s, for the same
total simulated time.

In each case the stress-strain curve shows a smooth regime
from zero strain up to a strain of about 0.03.  This smooth
response is indicative of elastic behavior in which the
lattice stretches but no dislocations are formed.  The
curves are lower than a straight-line extrapolation from
zero strain due to the non-linearity of the interatomic
potentials.  At some point the stress becomes large enough
that dislocations nucleate from the surface of the void.
The details of this process are described below.  The
dislocations leave a step on the void surface, expanding
the volume of the void and reducing the tensile stress
in the material.  This reduction takes place smoothly as
the dislocation loop is punched out and propagates away
from the void.  When the void growth rate is high enough
that the rate of increase of the void volume matches
and then exceeds the rate of increase of the simulation
box volume, the mean stress peaks and then decreases.  At
the highest strain rate, $\dot{\varepsilon}=10^9$/s, the
peak is rounded due to inertial effects.  Time is needed
for the release wave to propagate through the box so that
the stress reduction registers.  For a box size of
roughly 25 nm and a wave velocity of 5000 m/s, the wave transit
time is 5 ps, during which time the box strain increments by
0.005 at $\dot{\varepsilon}=10^9$/s.  So there is a smoothing
of the stress-strain curve for $\dot{\varepsilon}=10^9$/s
at this level.  The sound velocity also limits the
mechanisms of plastic relaxation, dislocation, twinning 
and fracture, also contributing to the rounded peaks.
At strain rates of $\dot{\varepsilon}=10^8$/s and less this
effect is much less significant, so the peaks are sharper.
At the lower strain rates, the peak stress is essentially
at the yield point; once the void surface has yielded,
the void is able to grow at a rate faster than the 
box is expanding.  At higher rates, a peak stress higher
than the yield stress is observed.  The yield and peak
stresses are given in Table \ref{table-threshold}.
For comparison, the yield point in copper was
found to be less, about 6~GPa \cite{SeppalaTriaxPRB}.

\begin{table}
\begin{center}
\begin{tabular}{|l|c|c|c|c|}
\hline
Metal &  $\dot{\varepsilon} =10^6$/s & 
$\dot{\varepsilon} =10^7$/s &
$\dot{\varepsilon} =10^8$/s &
$\dot{\varepsilon} =10^9$/s \\
\hline
V & & 16.48 & 17.24 (17.41) & 17.96 (19.11) \\
Nb & & 16.22 &16.41 (16.46)  & 17.65 (19.42) \\
Mo & & & 16.13 (16.17)  & 17.42 (18.13)  \\
Ta & 16.13 & 16.36 & 16.51 (17.47)  & 17.80 (18.99) \\
W & & 19.20 & 19.52 & 20.17 (20.66) \\
\hline
\end{tabular}
\caption{The void growth threshold stresses in GPa from
simulations of void growth in vanadium, niobium, molybdenum,
tantalum and tungsten at the strain rates indicated.
The peak stresses in GPa are shown in parentheses in cases 
where the peak is significantly higher than the threshold.
\label{table-threshold}
}
\end{center}
\end{table}

In some cases the stress-strain curve drops 
precipitously to a low value of stress; cf.\
tantalum $\dot{\varepsilon}=10^7$/s, $10^8$/s, 
and vanadium $\dot{\varepsilon}=10^8$/s.
In these cases the system twinned and
then cracked along a twin boundary.  The crack is able
to relieve a large amount of tension rapidly, accounting
for the precipitous drop in stress.

\subsection{Strain Rate Dependence}

Deformation processes such as void growth are strain rate
dependent if one or more of the mechanisms sets a time
scale.  In dislocation motion the time scale can be
set by a nucleation or multiplication process in which
some incubation time is needed to generate the dislocations
needed for plastic flow.  In some cases the rate dependence
arises because the mechanism is thermally activated, in
which case the process is temperature dependent as well,
and the use of a finite temperature molecular dynamics
simulation (as opposed to molecular statics) is important.

The stress-strain curves shown in Fig.\ 
\ref{fig-stress} have been calculated in
simulations with strain rates ranging from $10^6$/s
to $10^9$/s.  In each case the stress threshold for
void growth increases with increasing strain rate.  The
peak stress also increases with increasing strain rate,
as shown in Table \ref{table-threshold}.

\subsection{Comparison with Experiment}

Kanel and coworkers have studied the strain rate
dependence of the spall strength of Mo \cite{Kanel},
and it is interesting to compare their data to the
MD results, with the caveat that the comparison is not precise.
Their work used a flyer-plate and ablation drives to initiate spallation in
Mo single crystals, deformed single crystals, and polycrystals.
They considered single crystals aligned in different
high-symmetry orientations with respect to the wave.
They found that in each case the spall strength
increased with strain rate like $\sigma _{spall} \propto
\dot{\varepsilon}^\beta$ with an exponent $\beta \approx 0.30$.
The deformed single crystals and polycrystals exhibited
lower spall strengths than the undeformed single crystals.
The single crystal spall strength 
only had a weak dependence on orientation.  The exponent
of 0.3 corresponds to a two-fold increase in the spall strength for
every factor of 10 increase in the strain rate.

For the $<$100$>$ single crystal Mo, they found spall
strengths ranging from 3.3~GPa at $\dot{V}/V=3.6 \times 10^4$/s
to 13.53~GPa at $\dot{V}/V=3.0 \times 10^6$/s \cite{Kanel}, so
$\sigma _{spall} \approx 0.15~{\mathrm{GPa}} ~(\dot{V}/V)^{0.3}$.
Accounting for the relationship between the volumetric strain rate
and the linear strain rate, $\dot{V}/V=3 \dot{\varepsilon}$, 
this predicts a spall strength of $\sigma _{spall} = 53~{\mathrm{GPa}}$ 
at a strain rate of $\dot{\varepsilon}=10^8$/s.  The spall strength
includes a shear stress contribution that is absent from the
MD result.  The rate-dependent yield stress of Mo is not known.
It is expected to be significantly higher than the ambient yield
stress.  For comparison purposes, an upper limit is that the material 
does not yield prior to spall, so that the release wave is pure uniaxial
strain. This approximation, 
$\sigma _{mean} = (1 + 2 C_{12}/C_{11}) \sigma _{spall} /3$
where $C_{ij}$ are the Mo elastic constants, 
gives a mean stress at spall of 30~GPa
at $\dot{\varepsilon}=10^8$/s.  This extrapolation from the experimental
data is about twice the MD result for the peak stress at 
$\dot{\varepsilon}=10^8$/s (16.1 GPa at $\varepsilon=0.28$).  
This level of agreement with the experimental results may be reasonable
given the somewhat ill-defined nature of the spall strength, 
but it is not great.
Also the rate dependence is different: the MD results do not show a 
doubling of the peak stress when the strain rate is increased 10$\times$.
The strain-rate scaling for MD simulations of void growth in copper
is much closer to the Kanel scaling \cite{SeppalaTriaxPRB}.
%We will comment more about the comparison in the concluding section below.

\subsection{Void Volume}

\begin{figure}
\begin{center}
\resizebox*{90mm}{!}{\includegraphics{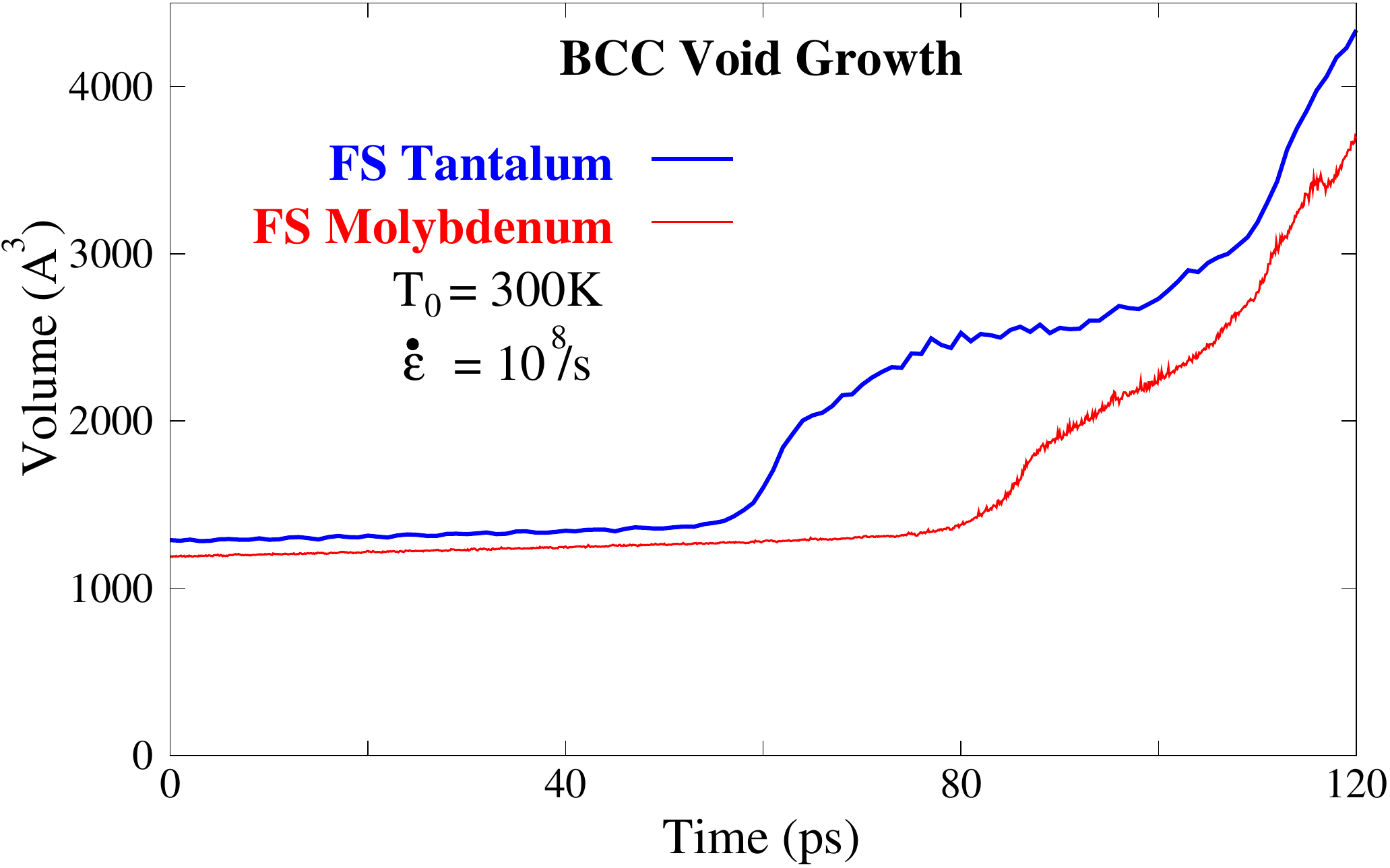}}%
\caption{Void volume as a function of time for Mo and Ta
at $\dot{\varepsilon}=10^8$/s.  The zero of time is offset
to correspond to strains $\varepsilon = 0.028$ and $0.030$
in Mo and Ta, respectively.
\label{fig-volume}
}
\end{center}
\end{figure}

Void growth is driven by the relaxation of the tensile stress
in the material surrounding the void.  The greater the volume
of the void, the less the matrix material needs to stretch
to fill the simulation box.  It is interesting to consider
how the void volume changes with time during the simulation.
We have developed techniques to detect the void surface
atoms and use them to calculate the volume of the void,
as well as other quantities that we do not consider here
such as the surface area and the shape as expressed in 
multipole moments \cite{DupuySurf}.  This analysis is
done on the fly as the simulation runs to reduce I/O
to the disk.  The results for the void volume as a 
function of time in Ta and Mo are shown in Fig.\ \ref{fig-volume}.

In the plot the curves begin with the elastic regime in which
the void is growing slightly as the metal stretches under
the applied tension.  Much of the elastic regime has taken
place before the offset zero of time in the plot.  Once
plasticity begins, the voids grow rapidly in bursts
separated by more quiescent periods.  This punctuated
growth corresponds to processes in the dislocation nucleation
that are explored below.  The relative magnitude of the steps in the
void growth is larger for these nanoscale voids that it would
be for larger voids since the ratio of the change in volume
from the emission of a dislocation loop to the void volume
goes like $b/R$, the ratio of the Burgers vector to the void
radius, the relative change is most important for nanoscale
voids.  

While the void volume is not currently measurable in dynamic
experiments, in principle the void size distribution could be
measured using small-angle x-ray scattering (SAXS) provided a sufficiently
intense x-ray source were available.  The Linear Coherent Light
Source (LCLS) and other fourth generation light sources may provide 
such the requisite intensity in the near future.
Thus far, a proof of principle SAXS measurement has been made 
in a static experiment on recovered, incipiently spalled titanium
at the Advanced Photon Source \cite{BelakSAXS}.
Also, tomography has been used to map the locations and sizes
of voids under static conditions also using recovered, incipiently
spalled samples \cite{BelakJCAMD}.  Because of the need for
x-rays from multiple lines of sight, tomography during dynamic
fracture is more challenging but not completely out of the question.

\section{Dislocations}
\label{sec-disl}

The material surrounding the void must deform plastically in
order for significant void growth to take place.  The 
traditional assumption in damage modelling is that the
stress surrounding the void increases to the yield strength
of the material and dislocations flow and multiply
according to the laws of continuum plasticity (see for example 
Ch.\ 5 of Hill \cite{hill}).  In this picture the plasticity
initiates in the vicinity of the void surface and the plastic
zone grows outward as the void grows.  Hill estimates the
size of the plastic zone after significant void growth
as $R_{plastic~zone}=(2E/3Y)^{1/3}R$, where $E$ is the
Young's modulus and $Y$ is the yield stress \cite{hill}.
This model assumes that there are sufficiently many dislocation
sources near the void so that material initially within a fraction 
of the void radius of the surface obeys continuum plasticity.

In modelling the high-rate damage and fracture processes, we
have taken a different view.  The initial void (or inclusion)
may be small compared to the length scale of the dislocation 
network.  In that case the prismatic dislocation flux needed
for the void growth does not flow from the matrix, but it
is emitted from the void surface.  In our single crystal
simulations, there are no dislocations initially, so they
must nucleate somewhere, and they are observed to nucleate
at the void surface in a process that is very similar 
to how loops were generated in the 
classic experiments of Mitchell \cite{Mitchell} in which hard
inclusions punched out dislocation loops in the transparent
silver chloride matrix due to stress build up as the temperature
changed.  The nucleation of dislocations from the void
surface requires higher stresses than typical yield stresses,
but as we saw above, the spall strength at high strain rates
is also very high so the driving stresses are available.

\begin{figure}
\begin{center}
\begin{minipage}{120mm}
\subfigure[]{
\resizebox*{40mm}{!}{\includegraphics{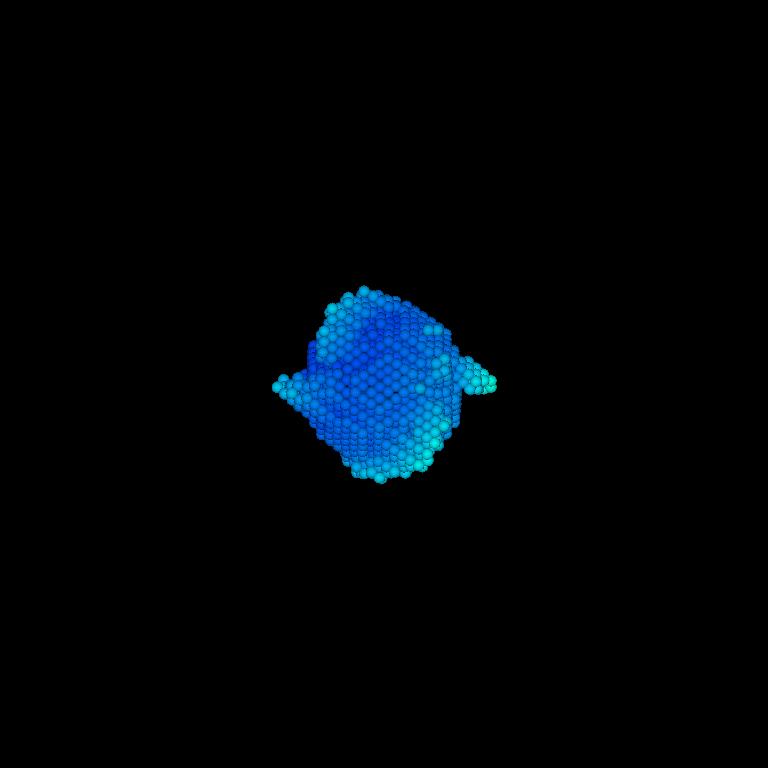}}}%
\subfigure[]{
\resizebox*{40mm}{!}{\includegraphics{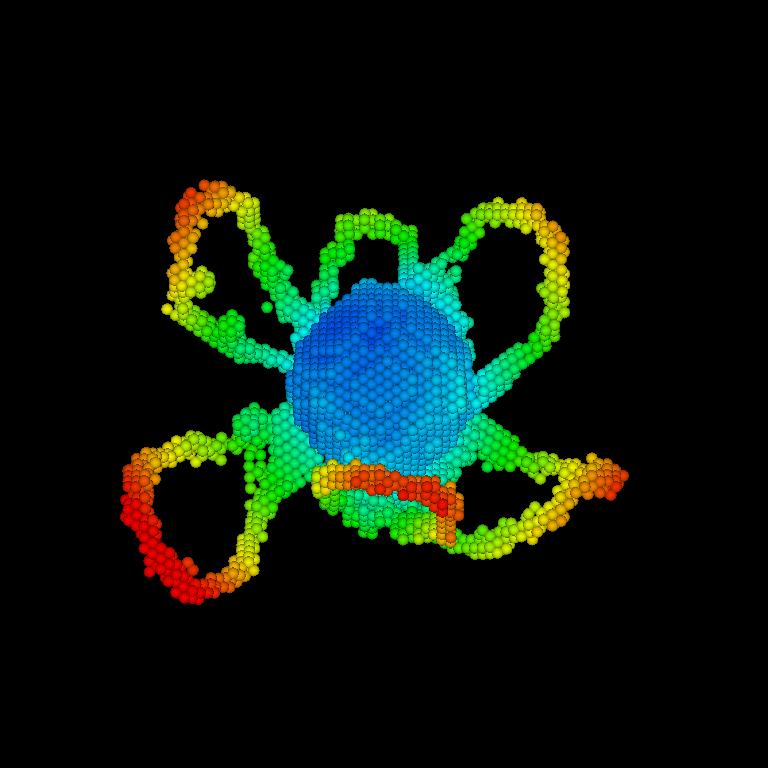}}}%
\subfigure[]{
\resizebox*{40mm}{!}{\includegraphics{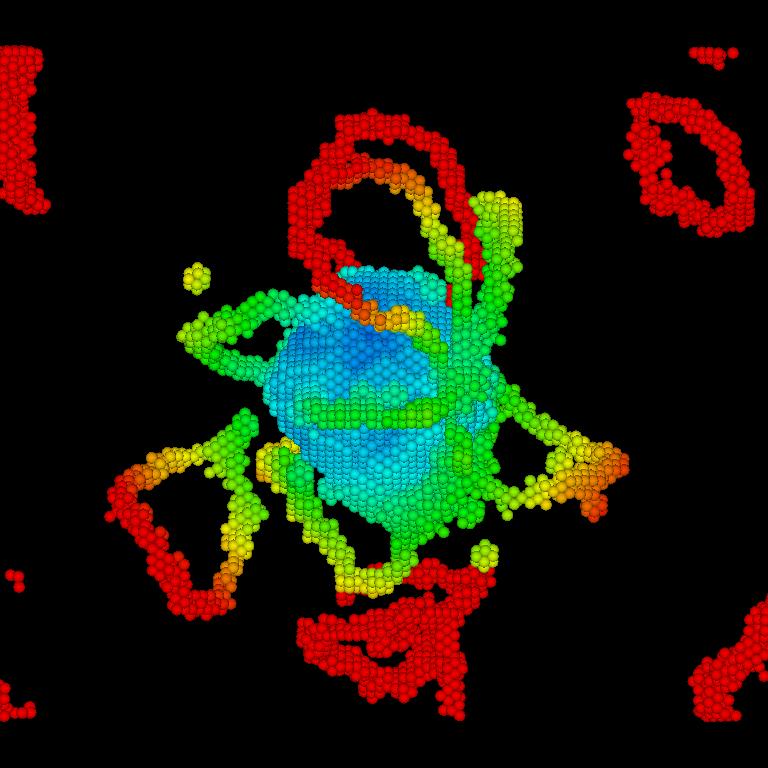}}}%
\caption{A time sequence of snapshots of dislocation
activity around a void growing in Mo at a strain rate of
$\dot{\varepsilon}=10^8$/s.
The snapshots are at times: (a) 80 ps, (b) 100 ps, and (c) 120 ps.  
The colouring indicates the distance from the centre of the void.
The void surface atoms are blue.
A small cluster of atoms to the upper left of the void consists
of atoms surrounding a vacancy created by dislocation climb.
\label{fig-disl-time}
}
\end{minipage}
\end{center}
\end{figure}

We have used the centrosymmetry deviation analysis to identify
dislocations that are generated as the material surrounding
the void deforms plastically to accommodate the growing void.
A series of snapshots of the dislocations in the material
surrounding a void growing in Mo is shown in Fig.\ \ref{fig-disl-time}.
As explained above, only atoms at defective lattice sites are
shown.  Dislocation loops consisting of full dislocations are
seen being punched out by the growing void.  
In Fig.\ \ref{fig-disl-time} each prismatic
loop consists of an edge dislocation surrounding a platelet
of interstitial atoms, effectively transporting those atoms
away from the void.  If the area of the loop is $A_{loop}$,
the volume of atoms is $b \, A_{loop}$, where $b$ is the
magnitude of the Burgers vector (2.73{\AA} in Mo).  This volume
is the amount by which the void grows from before the loop
emission once the loop is in the far field.  The loops
are prismatic in character here since the stress is essentially
hydrostatic.  Were a background shear stress present, shear
loops would be observed \cite{SeppalaTriaxPRB}, 
but it is still the prismatic component of the loops that 
leads to an increase in the void volume \cite{RuddJCAMD}.
Plastic flow prior to void growth limits the shear stress in
any case.

In Fig.\ \ref{fig-disl-time} a small cluster of atoms appears in the
visualisation to the upper left of the void.  This cluster consists
of atoms surrounding a vacancy created by dislocation climb.  The
cluster shows up in the visualisation because the atom missing
at the vacancy means one of the neighbours is poorly paired in
the centrosymmetry deviation, giving a value above the threshold.
The dislocation can climb to expand or reduce the size of the
loop, leaving a point defect behind.  Typically a few of these
defects
are observed in each void growth simulation.  This situation
is quite different from the massive amounts of debris formed
in the simulations of single dislocation glide reported 
in Ref.\ \cite{MarianBulatov}.
In that work approximately straight screw dislocations gliding in
a specific direction were observed to leave large clusters
of vacancies and interstitials behind. The debris formation
was attributed to cross-kink
formation.  Here the dislocations are edge dislocations, at least
once they are free of the void, and they are not gliding in
a special direction so there is no cross-kink competition.
Relatively little debris is formed.

\subsection{Prismatic Loop Geometry}

\begin{figure}
\begin{center}
\begin{minipage}{120mm}
\subfigure[]{
\resizebox*{40mm}{!}{\includegraphics{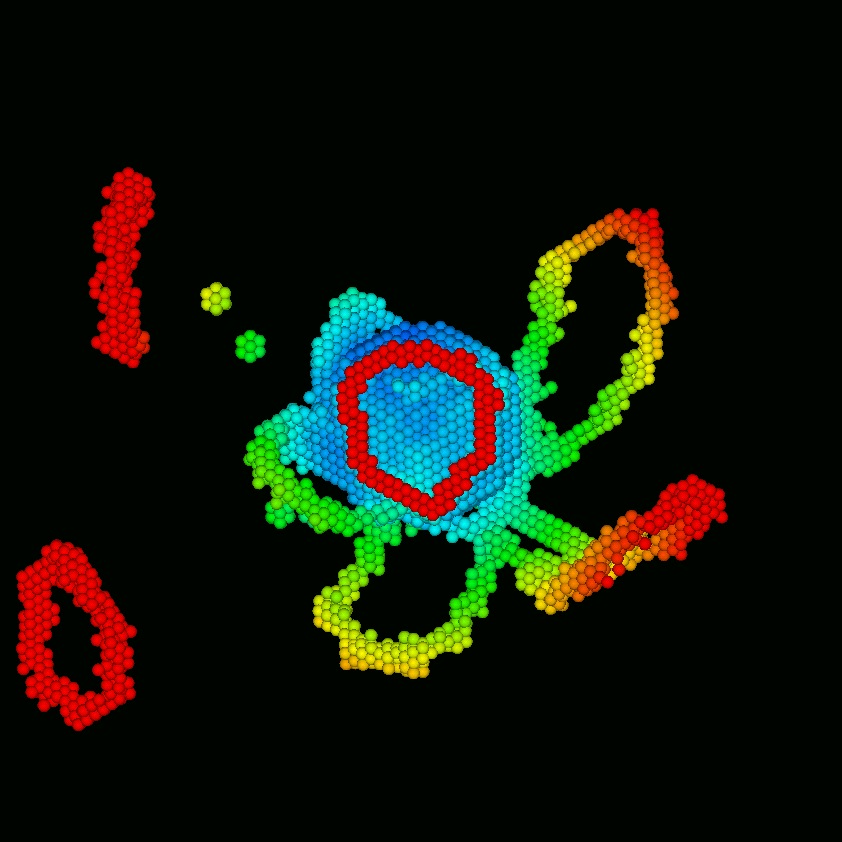}}}%
\subfigure[]{
\resizebox*{40mm}{!}{\includegraphics{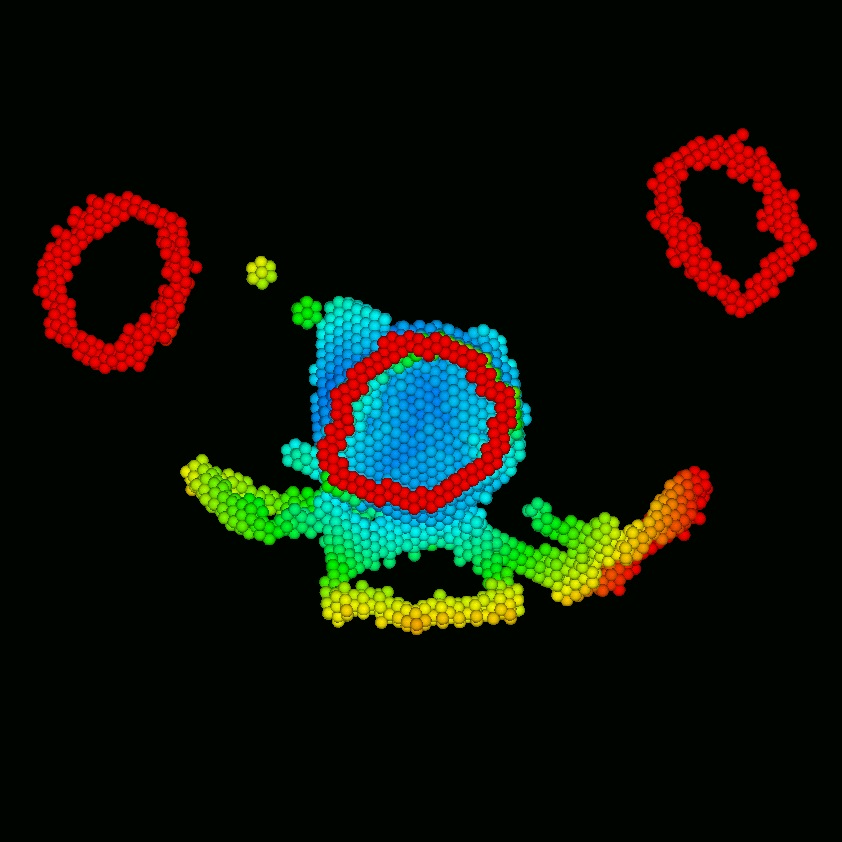}}}%
\subfigure[]{
\resizebox*{40mm}{!}{\includegraphics{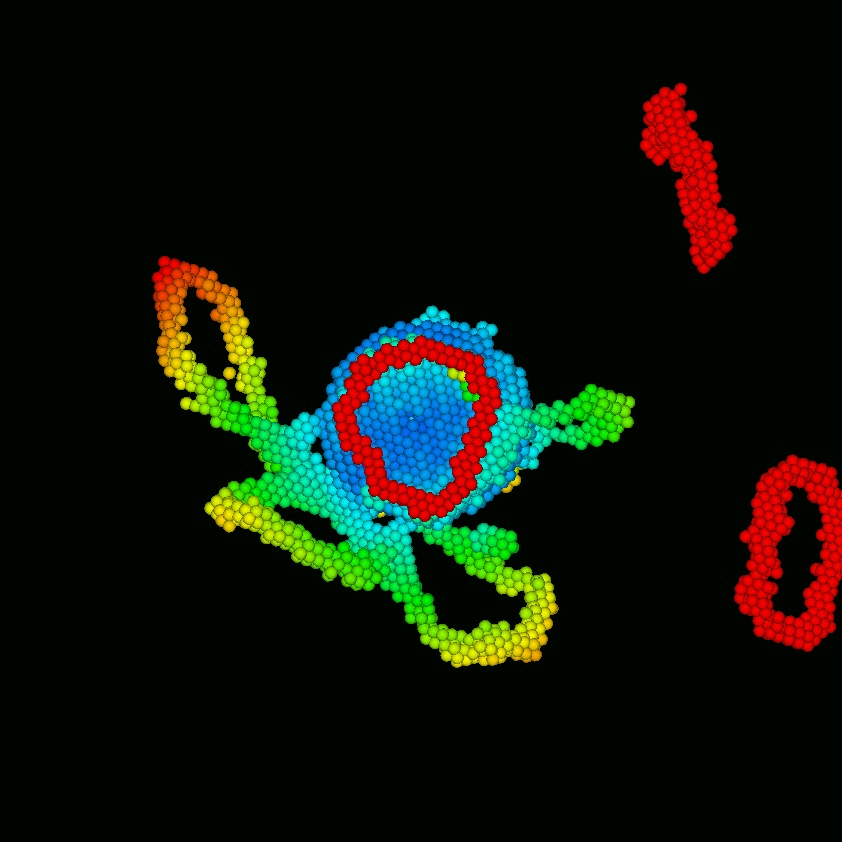}}}%
\caption{One configuration from the simulation of
void growth in Mo at a strain rate of
$\dot{\varepsilon}=10^8$/s viewed from three
different angles along $\langle 111 \rangle$ lines 
that are the glide directions of the prismatic loops.
The colouring indicates the distance from the centre of the void.
The void surface atoms are blue.
Two cluster of atoms consist
of atoms surrounding a vacancies created by dislocation climb.
\label{fig-disl-orient}
}
\end{minipage}
\end{center}
\end{figure}

Dislocation loops emitted from the void are shown 
in Fig.\ \ref{fig-disl-orient}.  The three panels of the figure 
show different views of the same dislocation configuration
punched out from a growing void, in each case looking down 
a different $\langle$111$\rangle$ line to the centre of the void in Mo.  
The Burgers vectors of the dislocations are 
$\frac{1}{2}\langle$111$\rangle$, 
as expected in bcc metals. 
It can be seen that the prismatic dislocation loops
are roughly circular in shape matching the spherical shape of
the void. Their size is a bit smaller than the void, and each
loop is propagating on a $\langle$111$\rangle$ glide cylinder
that is concentric with the void.  For a spherical inclusion or void 
in an elastically isotropic medium in pure hydrostatic tension
in the far field (cf.\ the stress field (\ref{eq-stress})), 
it is well known that the resolved shear stress at the surface of 
the void is greatest at $45^\circ$ from the centre line of the glide
cylinder \cite{HullBacon}.  A dislocation loop that nucleates at
this point of maximum resolved shear stress has a radius related
to that of the void
\begin{equation}
R_{loop} = R_{void}/\sqrt{2} .
\end{equation}
With the small voids it is difficult to test this relationship
very precisely, but it is approximately obeyed and the loops
are observed to increase in size as the void grows.

\subsection{Dislocation Nucleation}

\begin{figure}
\begin{center}
\begin{minipage}{100mm}
\subfigure[]{
\resizebox*{50mm}{!}{\includegraphics{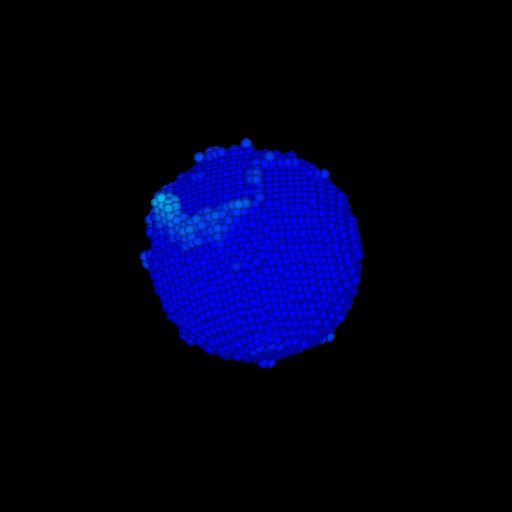}}}%
\subfigure[]{
\resizebox*{50mm}{!}{\includegraphics{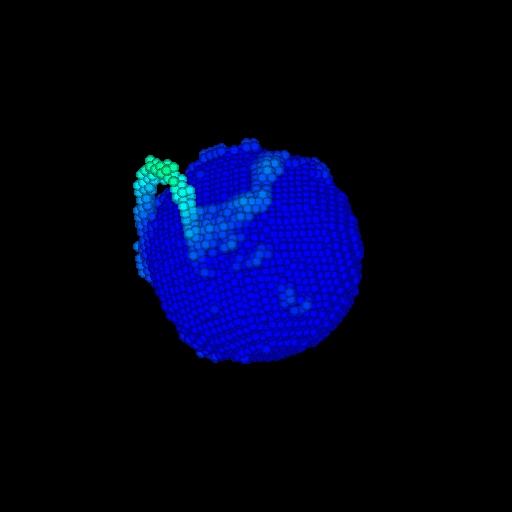}}}%
\\
\subfigure[]{
\resizebox*{50mm}{!}{\includegraphics{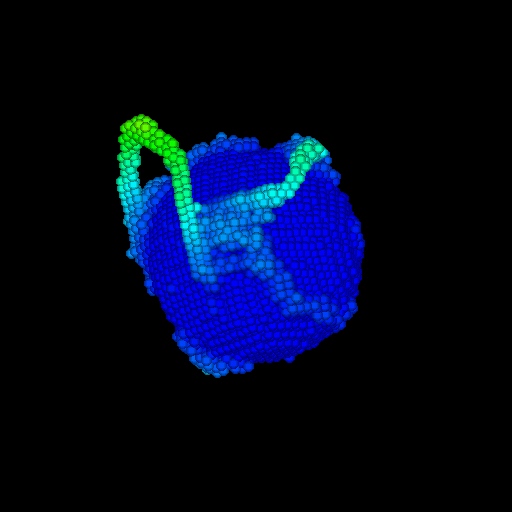}}}%
\subfigure[]{
\resizebox*{50mm}{!}{\includegraphics{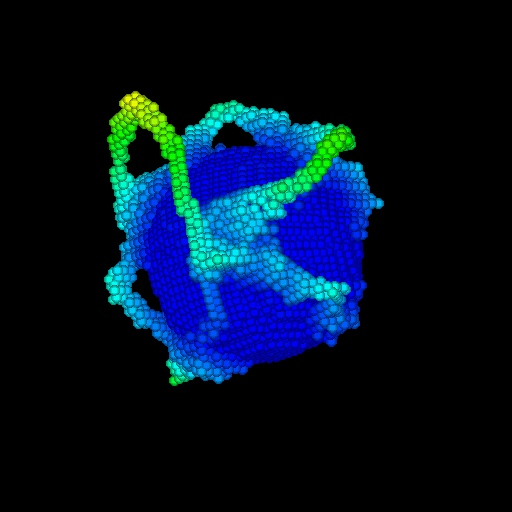}}}%
\caption{Dislocation nucleation from the surface of a 
void in a 2-million atom simulation of 
Ta at a strain rate of $\dot{\varepsilon}=10^8$/s.
The four panels show four different times during a
simulation of the growth process: (a) 7.5 ps, (b) 8.5 ps,
(c) 9.5 ps, and (d) 10.5 ps.  The colouring indicates
distance from the centre of the void.
\label{fig-disl-nucl}
}
\end{minipage}
\end{center}
\end{figure}

The mechanism of dislocation nucleation during void growth
in bcc metals is also interesting, and it differs in some
significant ways from the nucleation process in fcc metals.
For the purposes of this discussion we consider the nucleation
process to include everything from the beginning of dislocation-like
fluctuations on the surface of the void through the punch-out
process to the point that the prismatic loop has separated from
the void.  In many of our bcc simulations that nucleation 
process is well defined--once the leading segment of the loop
has glided a few void diameters from the void, the rest of the
loop has separated from the void and disentangled itself from
the other dislocations.  The initial stages of dislocation
punch out are shown in Fig.\ \ref{fig-disl-nucl}.
The first panel (Fig.\ \ref{fig-disl-nucl}a) shows a fluctuation
on the surface of the void as loop nucleation begins.
The second panel (Fig.\ \ref{fig-disl-nucl}b) shows a segment
of the dislocation loop extending from the surface of the void.
The other panels (Fig.\ \ref{fig-disl-nucl}c-d) show further
progression of the loop nucleation.  These snapshots are 
from a 2 million atom MD simulation of 
Ta void growth at a strain rate of $\dot{\varepsilon}=10^8$/s.
We have changed the simulation size for the nucleation study
since the 7~nm void in the larger simulation shows the details
of the nucleation process more clearly than the 4~nm voids
in the smaller simulations.

It is interesting to note that the dislocation configuration 
even in the early stages of nucleation does not exhibit cubic symmetry.
The configurations are observed to become more symmetric as
the strain rate is lowered and some metals (e.g.\ tungsten)
are more symmetric than others, but at all of the strain rates
explored here there is noticeable asymmetry.  Asymmetry can
result from the randomness of rare events, such as nucleation
on the cubically equivalent glide systems taking place through
thermally activated barrier crossings.  Averaged over time
the fluctuations appear to have a probability distribution
that respects cubic symmetry, but the actual nucleation events
break the symmetry. 

\begin{figure}
\begin{center}
\begin{minipage}{140mm}
\subfigure[]{
\resizebox*{70mm}{!}{\includegraphics{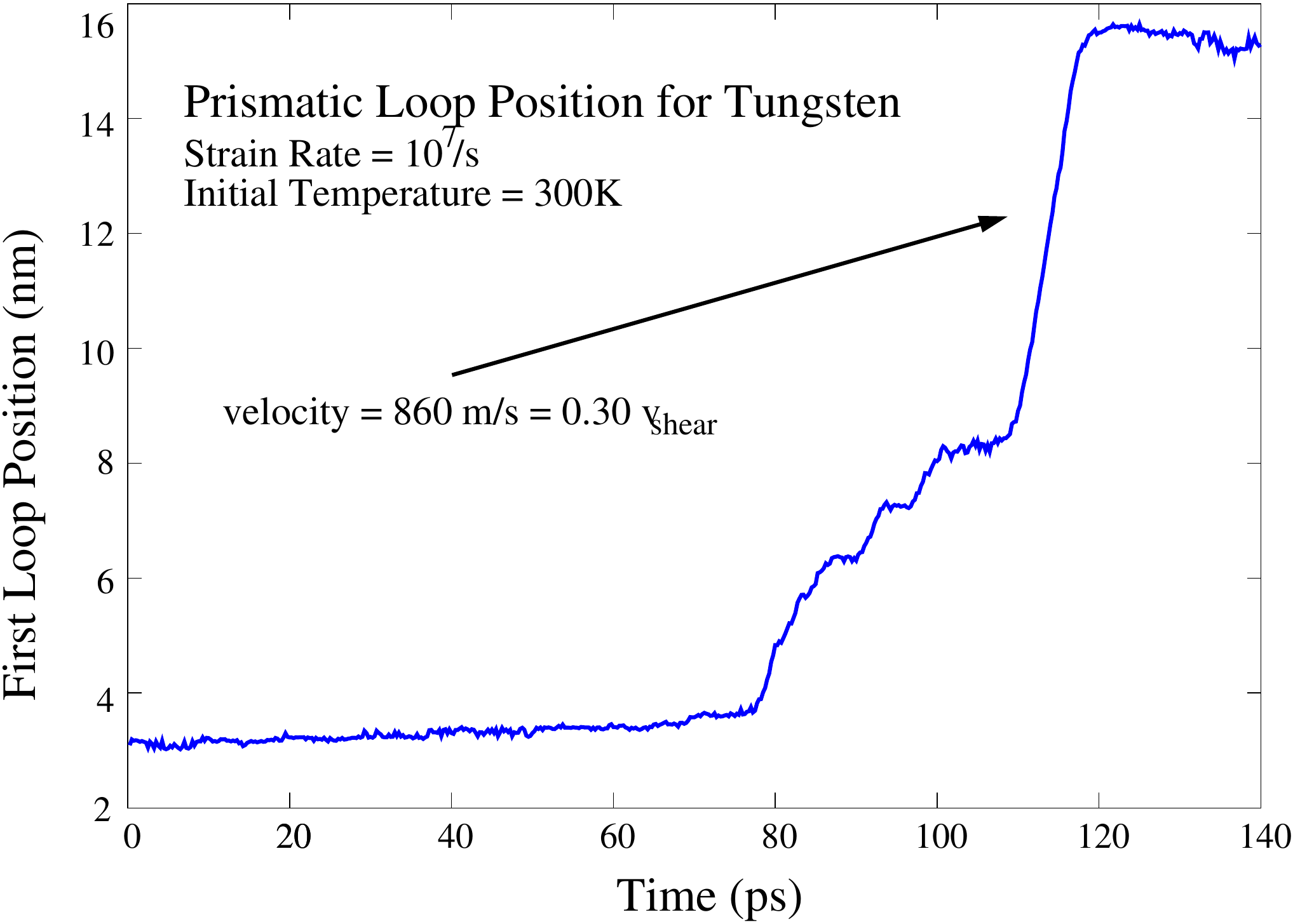}}}%
\subfigure[]{
\resizebox*{70mm}{!}{\includegraphics{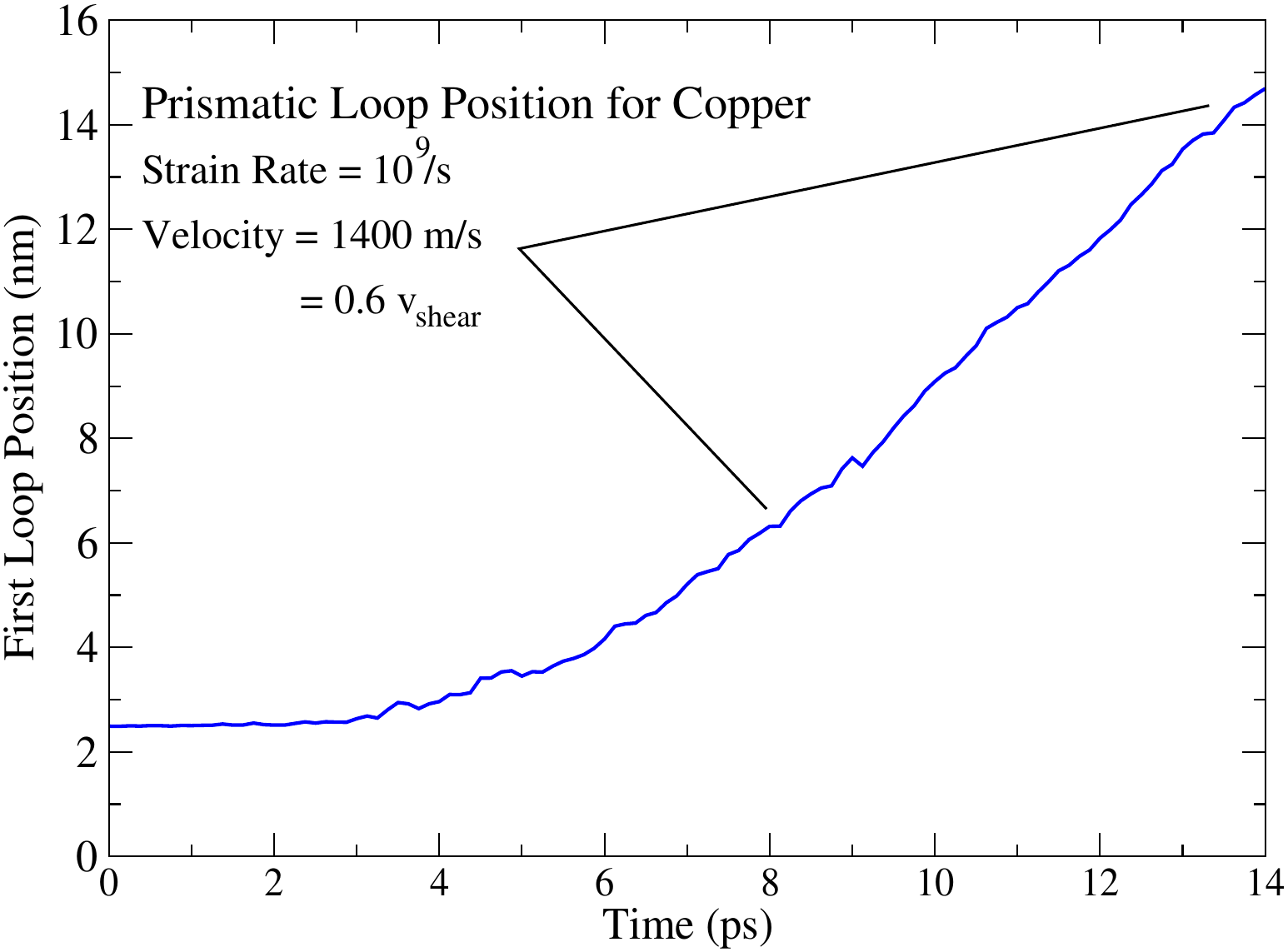}}}%
\caption{(a) Dislocation loop position as a function
of time in W at a strain rate of $\dot{\varepsilon}=10^7$/s.
The dislocation loop velocity in tungsten increases to 860 m/s, 
or about 30\% of the shear wave velocity. 
(b) Dislocation loop position as a function
of time in Cu at a strain rate of $\dot{\varepsilon}=10^9$/s.
The maximum loop velocity reached in this simulation was
1400 m/s, or about 60\% of
the shear wave velocity.
The loop velocity in the fcc metal is much smoother than
in the bcc metal.
}
\label{fig-disl-pos}
\end{minipage}
\end{center}
\end{figure}

In the bcc metals, the dislocation nucleation process
reaches a metastable configuration like that shown in
Fig.\ \ref{fig-disl-nucl}d, and it remains in that 
configuration for some time before continuing the 
nucleation process.  This quiescent period is apparent
in the plot of the position of the loop in 
Fig.\ \ref{fig-disl-pos}a from 100 ps to 105 ps. 
Here the loop position is taken to be the radial
distance of the leading edge of the loop from the
centre of the void.  
By contrast, the position of the loop in the fcc metal \cite{Moriarty2002}
shown in 
Fig.\ \ref{fig-disl-pos}b is much smoother in time.
In both plots the zero time has been offset by an
arbitrary amount.  The plateau in the position of
the loop in W at $\sim$120~ps (Fig.\ \ref{fig-disl-pos}a) is due to
the loop reaching the boundary of the simulation
box where symmetry dictates the driving stress goes
to zero so the loop stops moving.

The basic mechanism of this process can be 
understood by examining its energetics, 
involving the relaxation of the strain energy as the loop
is driven by the stress field of the void.  Also important 
is the line energy of the dislocation and the lattice
energy (Peierls barrier) that depends on where the 
dislocation core is located with respect to the
lattice.  For our purposes, we will neglect the 
change in the surface area of the void and the energy
associated with elastic image forces.
The total energy is then
\begin{equation}
E = E_{elastic} + E_{line} + E_{lattice}.
\end{equation}
Each of these terms can be approximated.  
The Peach-Kohler force per unit dislocation length is given by
${\mathbf{F}}_{PK}/L=\left( {\mathbf{b}}\cdot \sigma \right) \times {\mathbf{\xi}}$
where ${\mathbf{b}}$ is the Burgers vector and 
${\mathbf{\xi}}$ is the unit vector 
in the line direction \cite{HirthLothe}.
The Peach-Kohler force is projected onto the glide direction 
$\hat{\mathbf{n}}_{glide}$.
We use the void stress field $\sigma$ (\ref{eq-stress}), and
$E_{line} = \int \, dl\, \int \, dr \, {\mathbf{F}}_{PK}\cdot \hat{\mathbf{n}}_{glide}/L$.
The line energy per unit length is given by $E_{line}/L=\alpha G b^2$,
where $G$ is the shear modulus and $\alpha$ is a material-dependent 
constant of order unity \cite{HullBacon}.  We multiply this by the 
loop length, either $2\pi R_{loop}$ for a detached loop or 
$\pi R_{loop} + 2 (r-R_{void})$ for a loop that is still
attached to the void surface.  The final term is the 
Peierls energy $E_{lattice}=U \sin\left(2\pi r/b \right) L_{lead}$
where $U$ is the height of the Peierls barrier \cite{HullBacon} 
and $L_{lead}$ is the length of the leading edge of the dislocation, 
either $2\pi R_{loop}$ (detached) or $\pi R_{loop}$ (attached).

\begin{figure}
\begin{center}
\resizebox*{90mm}{!}{\includegraphics{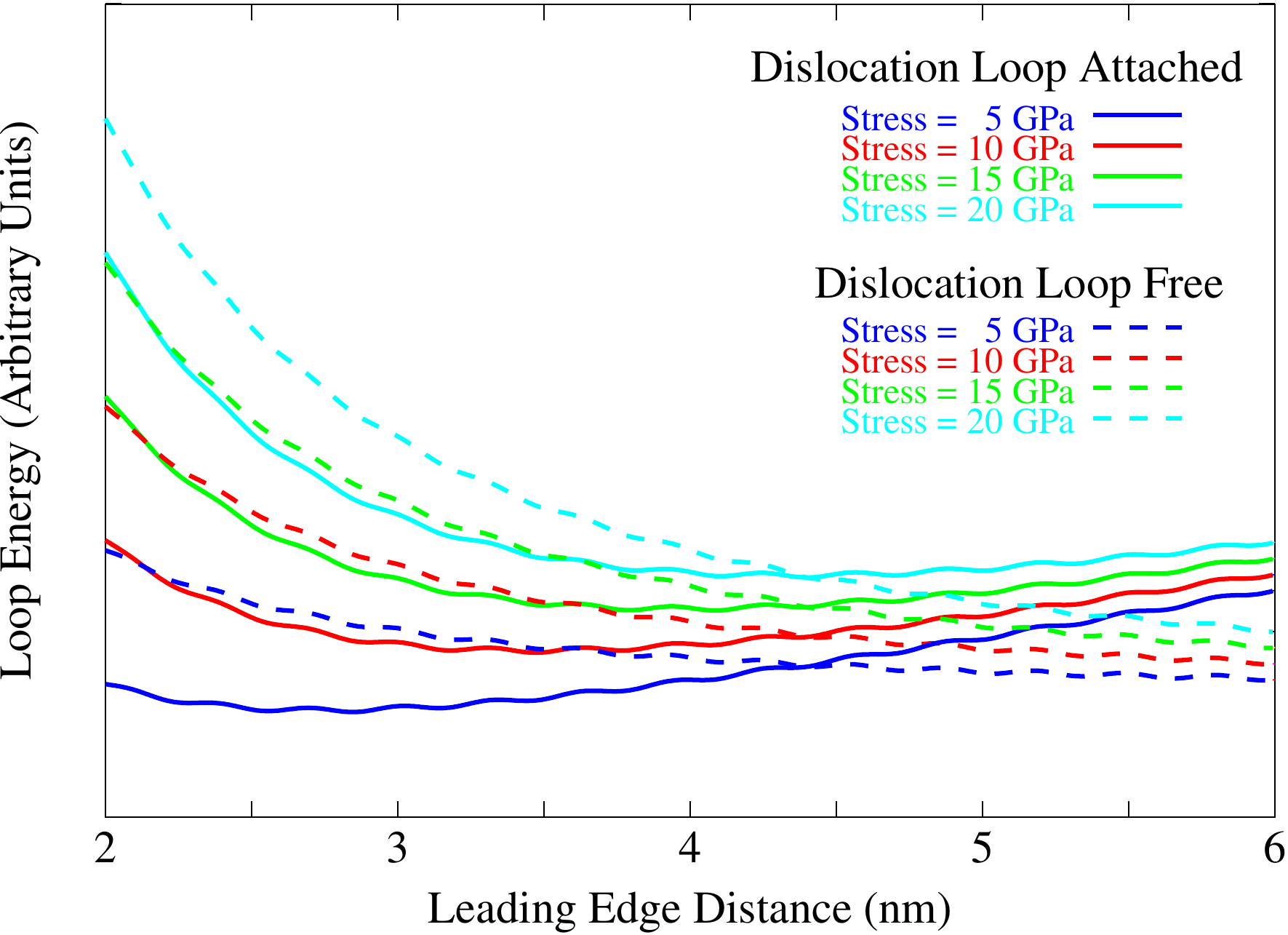}}%
\caption{Energetics of the dislocation nucleation process 
in a bcc metal, showing energy as a function of the
position of the leading edge of the loop for attached
and free loops.  The dislocation energy is approximated
as a sum of elastic, line tension and Peierls energy
contributions.  Initially it is energetically 
favourable for part of the loop to nucleation and
extend out from the void.  As the stress is increased,
the loops extends further out until the point the
free loop energy drops below that of the attached
loop, and it becomes favourable to nucleate the
full loop.
\label{fig-loopEnergetics}
}
\end{center}
\end{figure}

Using approximate parameters for tantalum, the 
resulting energetics of attached and detached loops
are plotted for different tensile stresses in
Fig.\ \ref{fig-loopEnergetics}.
The plots show that at lower tensile stresses the leading edge 
of the dislocation is driven by the elastic energy to 
extend out from the void, as the Peach-Kohler forces overcome
the line tension.  As the dislocation moves outward, the shear stress drops
and the upward turning energy and the lattice barrier stop the
dislocation from moving further.  As the stress increases,
the dislocation moves further out.  Eventually,
it becomes advantageous to nucleate the rest of the loop
so that the detached loop can propagate further from the
void, not confined by the line tension.
It is interesting to note that the size of the
plastic zone in the initially dislocation free
crystal is set by the lattice resistance (Peierls
barrier) rather than a yield stress due to the
dislocation network.

In the fcc metals, the loop nucleation process is much smoother 
because the Peierls stress is very low compared to the stress 
needed to nucleate the dislocation from the void surface.  Once 
the loop is punched out, the Peierls barrier provides little 
resistance to the further glide of the loop.  In the bcc metals
the Peierls barrier is much higher.  It is highest for screw
dislocations, but the edge dislocations in the prismatic loops
of interest here also have a high Peierls stress, albeit not has high as for
screw dislocations \cite{Hirsch1960}.

To date there is little experimental evidence for a plastic
zone around a void after it has grown, and no means of
imaging individual dislocations around a void.  One transmission
electron micrograph that suggests a plastic zone around a void 
has been presented
by Meyers \cite{MeyersBook} without quantitative analysis.  
A differential shading around a void in aluminium sample
that was sliced and etched has been suggested to be
due to dislocation etch pits in the plastic zone \cite{SofronisVoid}.
At this point it is not expected that it will be possible
to image dislocations around a void {\em in situ} during
dynamic fracture any time soon, but improved techniques for
quantifying the plastic zone around voids in recovered
samples are likely.

\subsection{Twinning and Intergranular Fracture}

In many metals at high deformation rates or low temperatures,
dislocation-mediated plastic deformation cannot support the shear strain rate
so the shear stress rises until it reaches a threshold
at which twinning occurs \cite{Christian}.  Twins propagate rapidly,
leading to a plastic strain rate that equals the rate of
change of the twinned volume fraction times the eigenstrain
of the twin.  A similar phenomenon is observed to occur
in these simulations of bcc void growth.  If the void growth
rate, limited by the rate of nucleation and propagation of
dislocations, is too slow, the tensile stress will continue
to rise.  The stress field around the void changes the tensile
stress into a shear stress, and once the shear stress reaches
the twinning threshold, twins nucleate and propagate rapidly.
If the void grows fast enough, the threshold is never reached.

\begin{figure}
\begin{center}
\resizebox*{70mm}{!}{\includegraphics{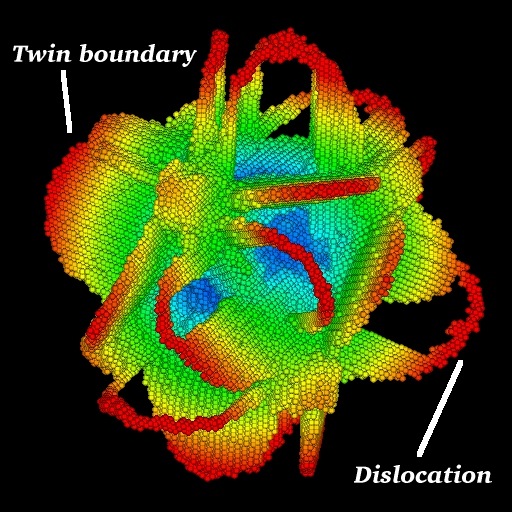}}%
\caption{Twin nucleation at the surface of a void in
Ta at a strain rate of $\dot{\varepsilon}=10^9$/s.
The colouring indicates the distance from the centre of
the void.  The sheets of atoms in the figure are
twin boundaries; some dislocations are present, too.
\label{fig-twin-nucleation}
}
\end{center}
\end{figure}

In practice we observe a threshold in the strain rate such that
for sufficiently high strain rates, twins form at the void surface
and propagate through the simulation box.  The twin boundaries
are then observed to fracture leading to a rapid drop in the
tensile stress.  A visualisation of twin nucleation at the
void surface in Ta at a strain rate of $\dot{\varepsilon}=10^9$/s
is shown in Fig.\ \ref{fig-twin-nucleation}.  

\begin{figure}
\begin{center}
\begin{minipage}{140mm}
\subfigure[]{
\resizebox*{70mm}{!}{\includegraphics{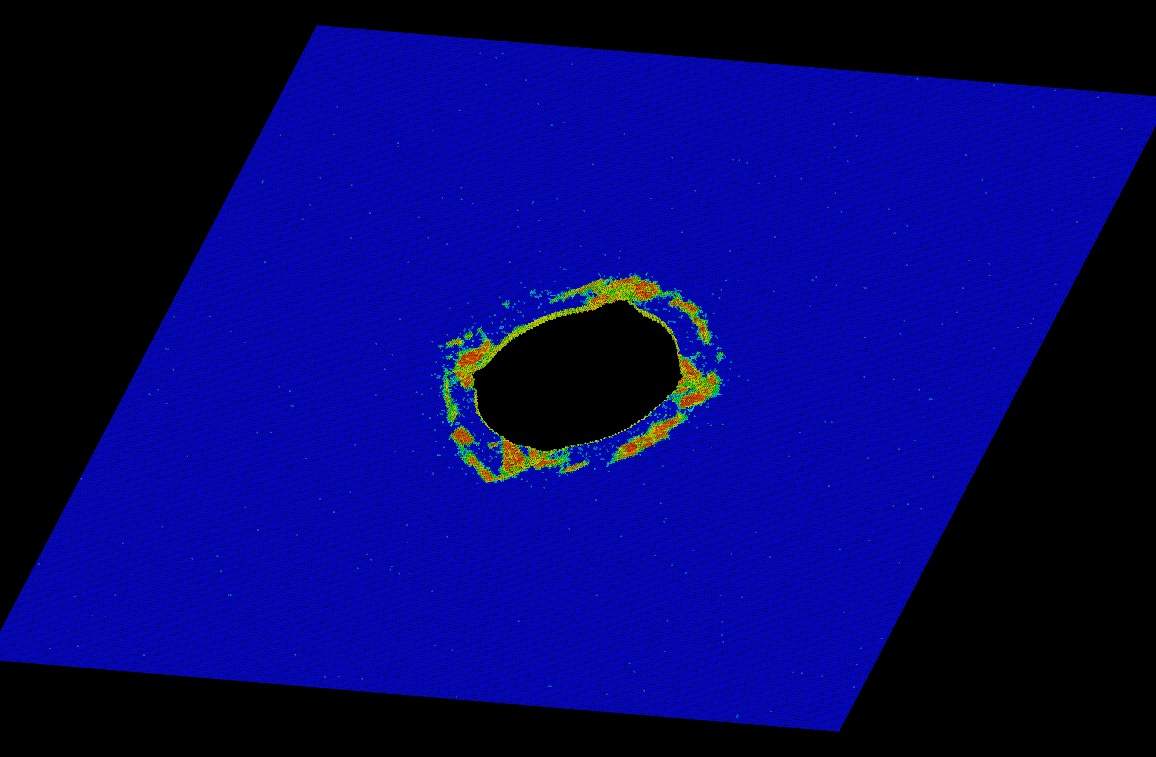}}}%
\subfigure[]{
\resizebox*{70mm}{!}{\includegraphics{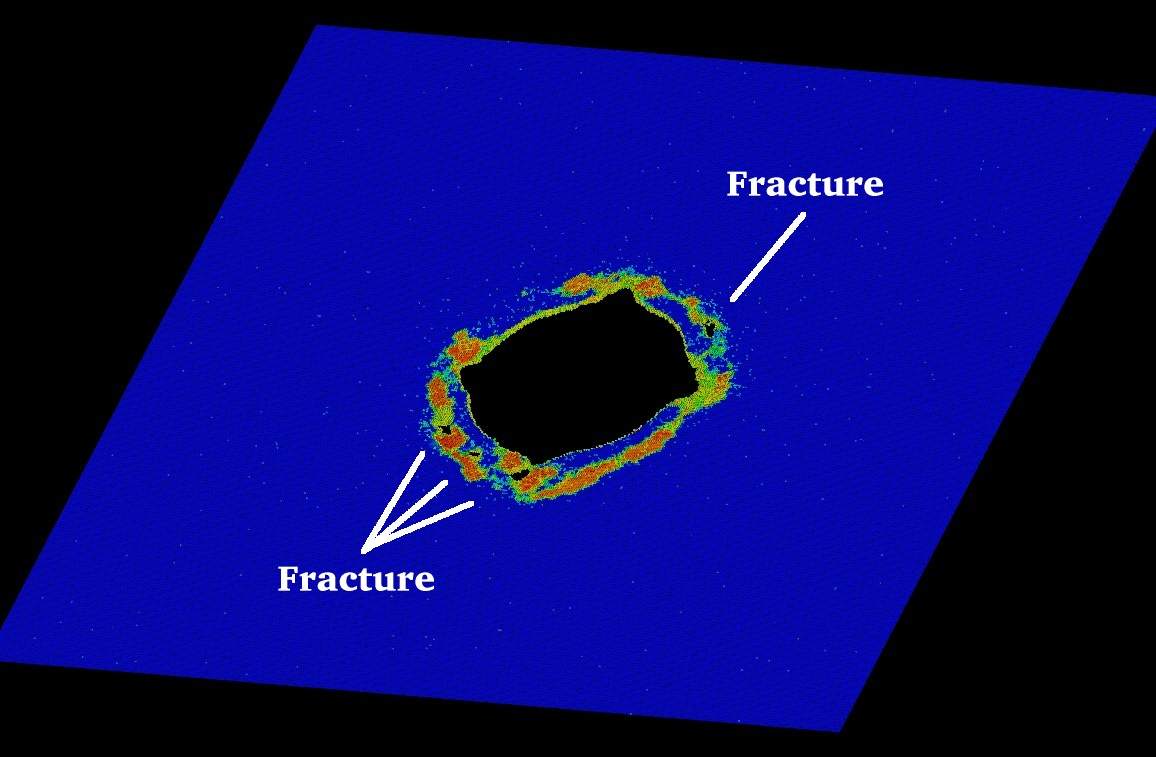}}}%
\\
\subfigure[]{
\resizebox*{70mm}{!}{\includegraphics{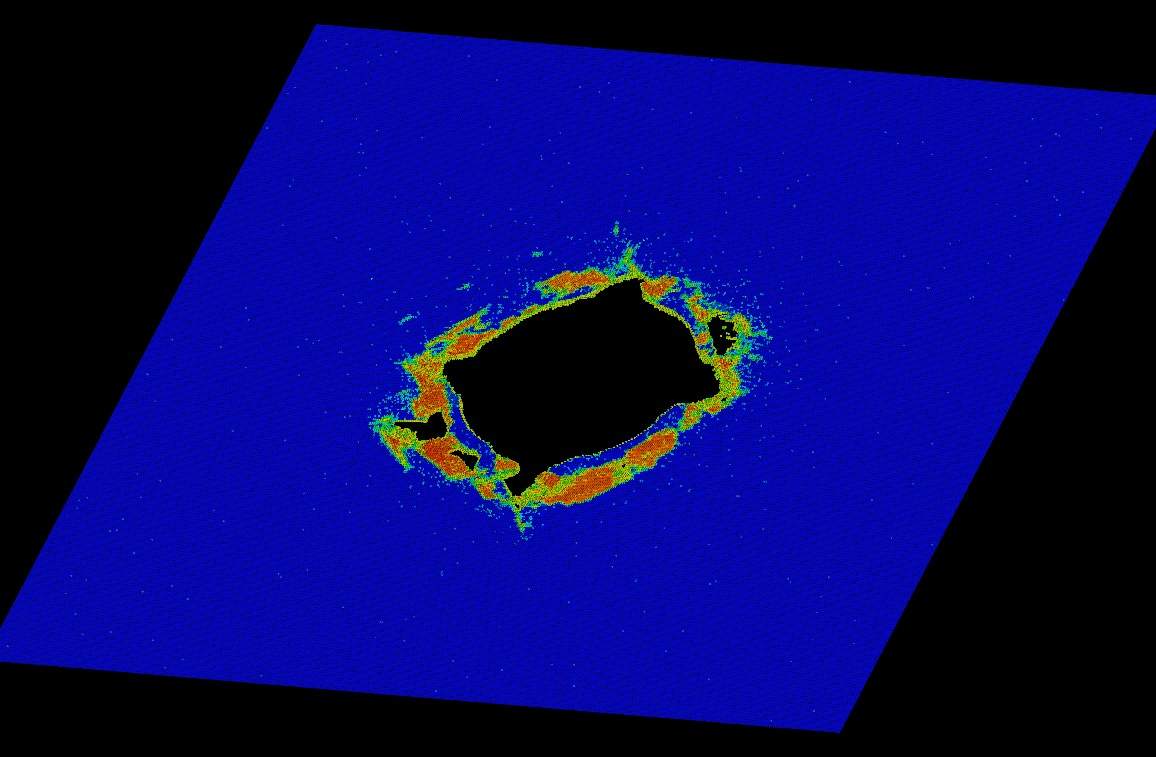}}}%
\subfigure[]{
\resizebox*{70mm}{!}{\includegraphics{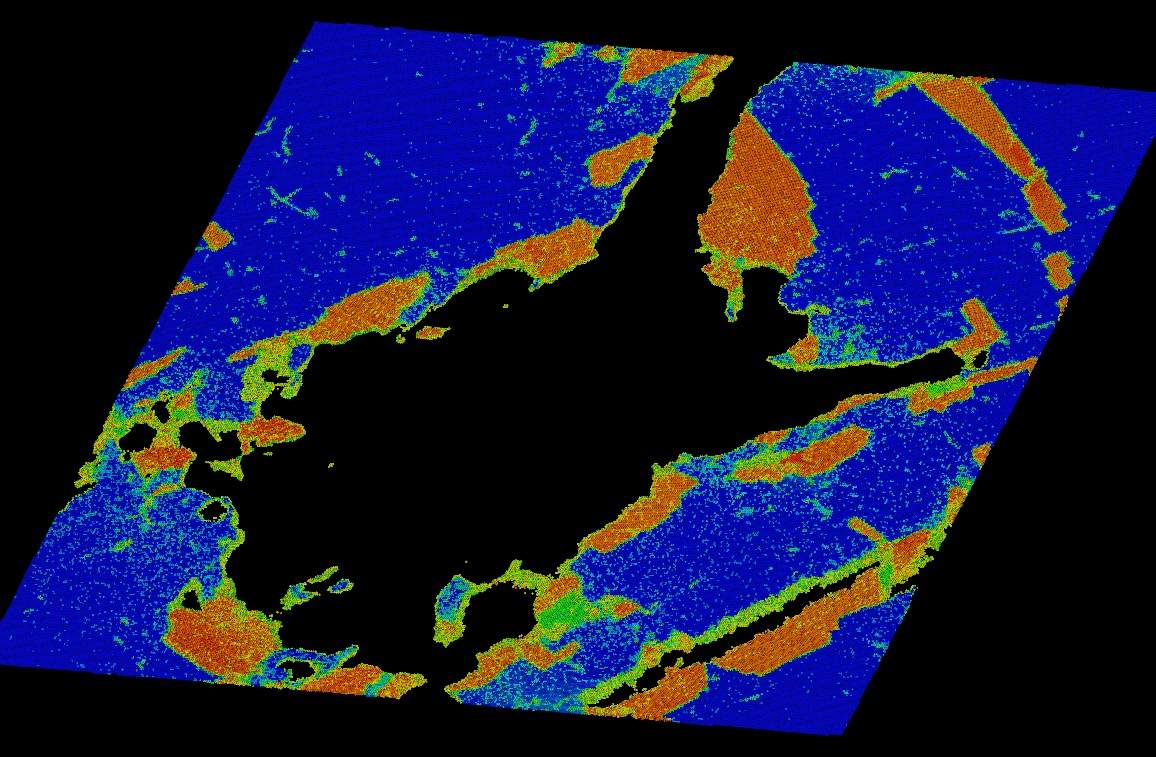}}}%
\caption{Orientation pattern around a void in
Mo at a strain rate of $\dot{\varepsilon}=10^8$/s in
a 128 million atom simulation.
The snapshot sequence shows 
(a) twinning at $\varepsilon = 0.0213$,
(b) more twinning and the start of fracture at $\varepsilon = 0.0228$,
(c) twinning and more developed fracture at $\varepsilon = 0.0244$,
and
(d) fracture spanning the simulation box at $\varepsilon = 0.0322$.
The atoms in a slice through the simulation box are shown.
The colouring indicates the local orientation of the lattice.
\label{fig-twinned}
}
\end{minipage}
\end{center}
\end{figure}

A series of
snapshots of a slice through a system as it twins and fractures 
is shown in Fig.\ \ref{fig-twinned},
coloured according to the lattice orientation.
The change in orientation due to the twins can be
seen near the void surface in Fig.\ \ref{fig-twinned}a.
The apparent ring of twinned material around the void
is due to twins propagating at an angle from the
surface as seen in Fig.\ \ref{fig-twin-nucleation}.
In the second panel, Fig.\ \ref{fig-twinned}b, fracture
is beginning to take place along the twin boundaries,
and it progresses further in Fig.\ \ref{fig-twinned}c.
The first three panels are spaced equally in time (and strain).
The fourth panel is later in time after the system has
completely fractured.  The time at which the spanning
fracture occurs depends on system size, and in a real
material fractures like this would link up between voids
in a more random fashion rather than immediately spanning
the system, but the basic process of twin-mediated fracture
is plausible at high rates.

\section{Comparison with FCC Voids}
\label{sec-fcc}

The basic mechanism of void growth is the same in
bcc and fcc metals, but there are significant
differences in some aspects of the process.  The Peierls
barrier is much lower in fcc metals than in bcc metals,
and this difference accounts for many of the differences
in the dislocation processes.  Also, in fcc metals the
stacking fault energy is relatively low so the dislocations
are split into partials \cite{HullBacon}. These partials
limit cross-slip and tend to produce dislocations that
are linear over persistence lengths of several nanometres
or more.  As a result, the prismatic dislocation loops from nanoscale
voids form as parallelograms with four straight legs on the
\{111\} glide planes (cf.\ Fig.\ \ref{fig-vis}).  The loops
in bcc metals, on the other hand, are formed from perfect dislocations
rather than partials spanned by stacking fault ribbons, and they
have a shorter persistence length, so that loops are more circular
even at the nanoscale.

In copper the void surface evolved from an initially spherical
shape to a more faceted shape, with low energy \{111\} facets.  The edges
connected the facets were rounded, but the void had a definite
octahedroid shape, as quantified by the hexadecapole moment
\cite{SeppalaTriaxPRB}.  In the bcc metals studied here
there is no pronounced faceting of the voids.  The \{110\}
surfaces are the close-packed surfaces of the bcc metals, but
they do not appear to play a special role on the bcc void
surfaces.  The voids do not take on the shape of a dodecahedron,
although if the edges are rounded, a dodecahedron is very
similar to a sphere and it may be that the voids here are
too small to see the effect.

In bcc metals the motion of the loops is different, too.  
The loop nucleation
process is fairly smooth in fcc metals, whereas in bcc metals
the initial motion of the is jerky as shown in Fig.\ \ref{fig-disl-pos}
for tungsten and copper.
In both cases the velocities of prismatic loops can be a 
large fraction of the shear wave velocity.  The maximum
velocities in the figure are (a) 860 m/s = 30\% of the shear wave velocity
for tungsten and (b) 1400 m/s = 60\% of the shear wave velocity
for copper.

\section{Nucleation in Polycrystals}
\label{sec-polycrystal}

The process of nucleation of voids at grain boundaries in 
polycrystalline systems is also interesting, both from the
point of view of how grain boundaries affect fracture
and from the point of view of the extraordinary mechanical
properties of nanocrystalline metals \cite{Wolf,HvS}.  
Nanocrystalline tantalum, in particular, has been 
considered for applications in inertially confined
fusion where it would be subjected to high rate loading
\cite{Hamza}.
We have investigated dynamic fracture
in these systems through a series of simulations of void
nucleation and growth in an initially fully dense 16-million-atom
MD simulation of nanocrystalline Ta.  These systems were 
produced in previous work by Streitz et al.\ \cite{StreitzSolid}
through rapid compression of molten Ta. Previously, we have
investigated the plastic flow in this system under uniaxial
and biaxial strain at constant volume \cite{RuddMSF}.  Here
we investigate how the system responds through the nucleation,
growth and coalescence of voids as the volume is
increased at a specified strain rate.

\begin{figure}
\begin{center}
\begin{minipage}{100mm}
\resizebox*{100mm}{!}{\includegraphics{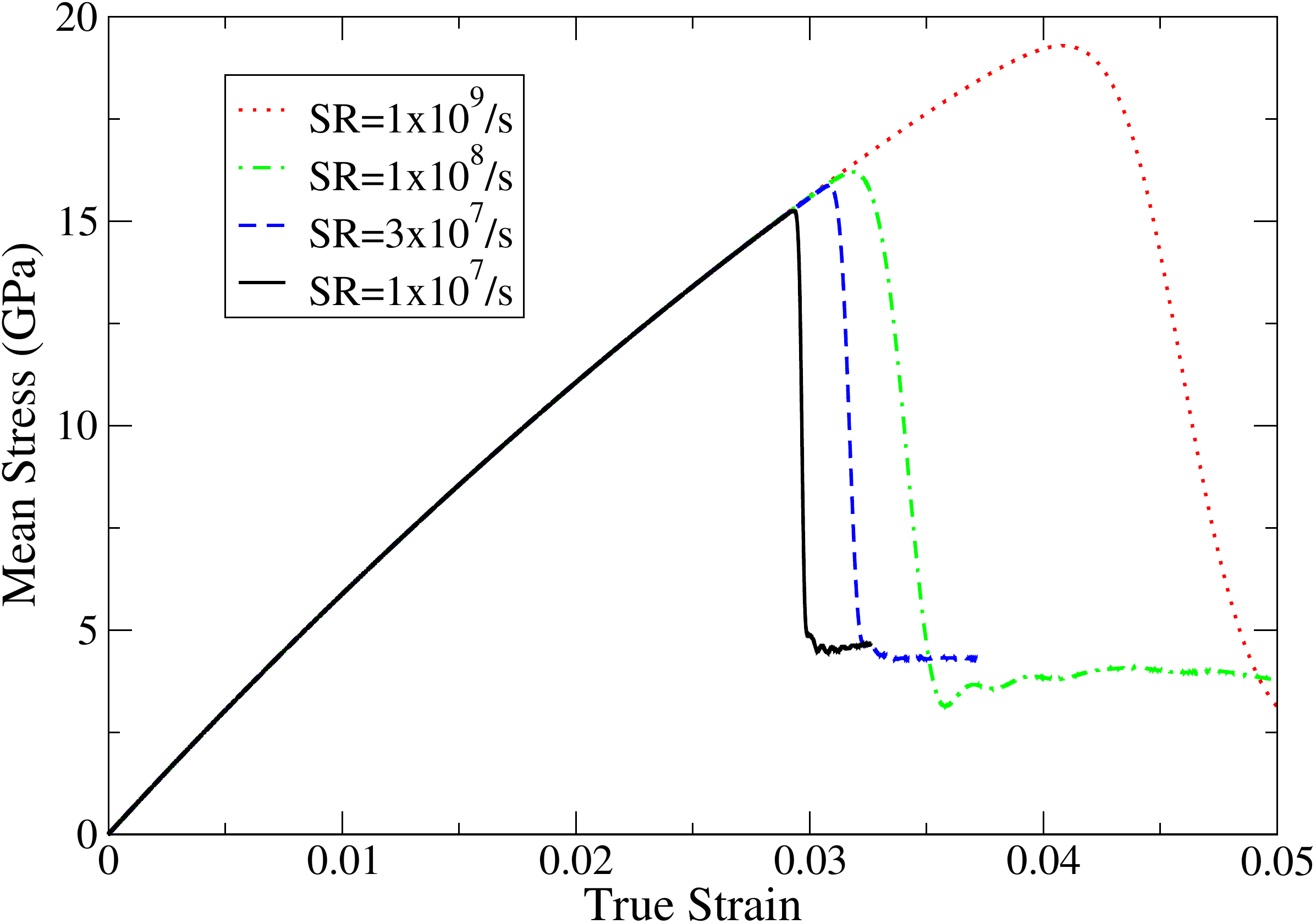}}%
\caption{Stress-strain curve for a nanocrystalline Ta system undergoing
expansion at the specified strain rates.
All of the simulations began at room temperature.  
The mean stress, plotted on the vertical axis, 
is the negative of the pressure.
}
\label{fig-nanoxtal-stress-strain}
\end{minipage}
\end{center}
\end{figure}

Figure \ref{fig-nanoxtal-stress-strain} shows the stress-strain
curve for a nanocrystalline system undergoing dynamic fracture
as simulated in molecular dynamics for several different
strain rates.  As in the single crystal simulations,
there is an elastic phase followed by
the onset of plasticity and substantial void volume increase
leading to a drop in the stress.  Also like the single crystal simulations,
the peak stress increases as the strain rate increases.

\begin{figure}
\begin{center}
\begin{minipage}{140mm}
\subfigure[]{
\resizebox*{70mm}{!}{\includegraphics{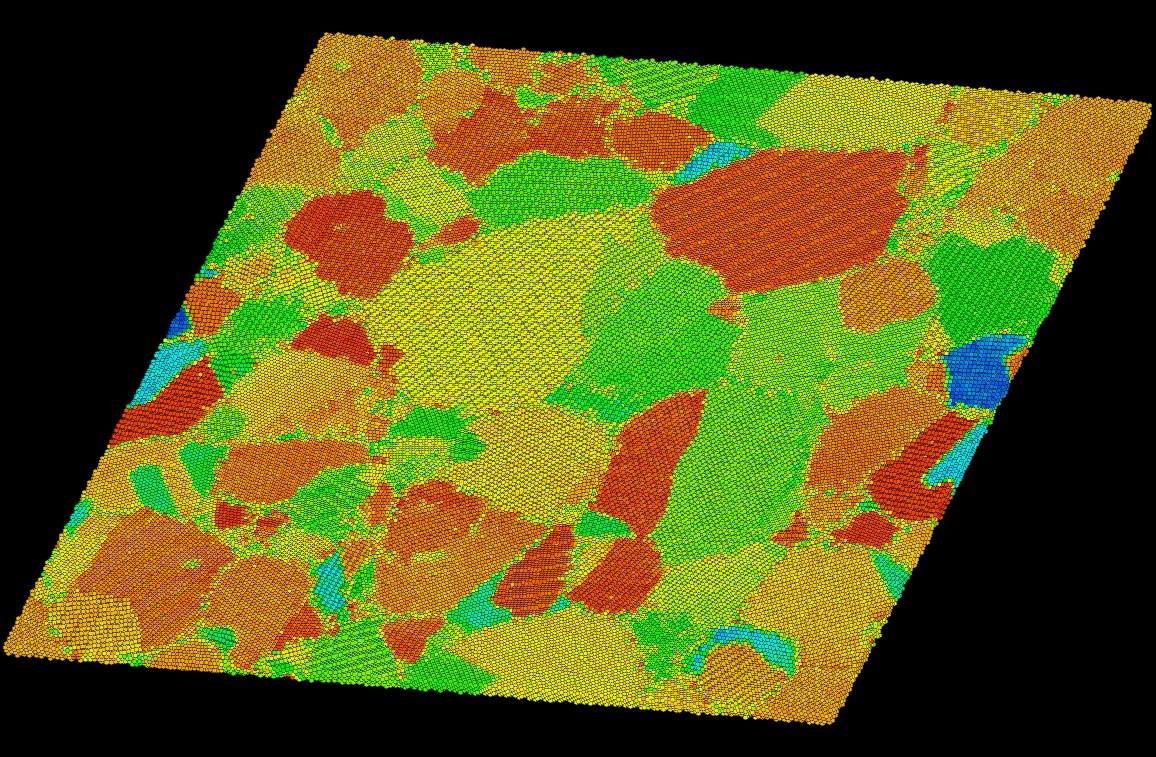}}}%
\subfigure[]{
\resizebox*{70mm}{!}{\includegraphics{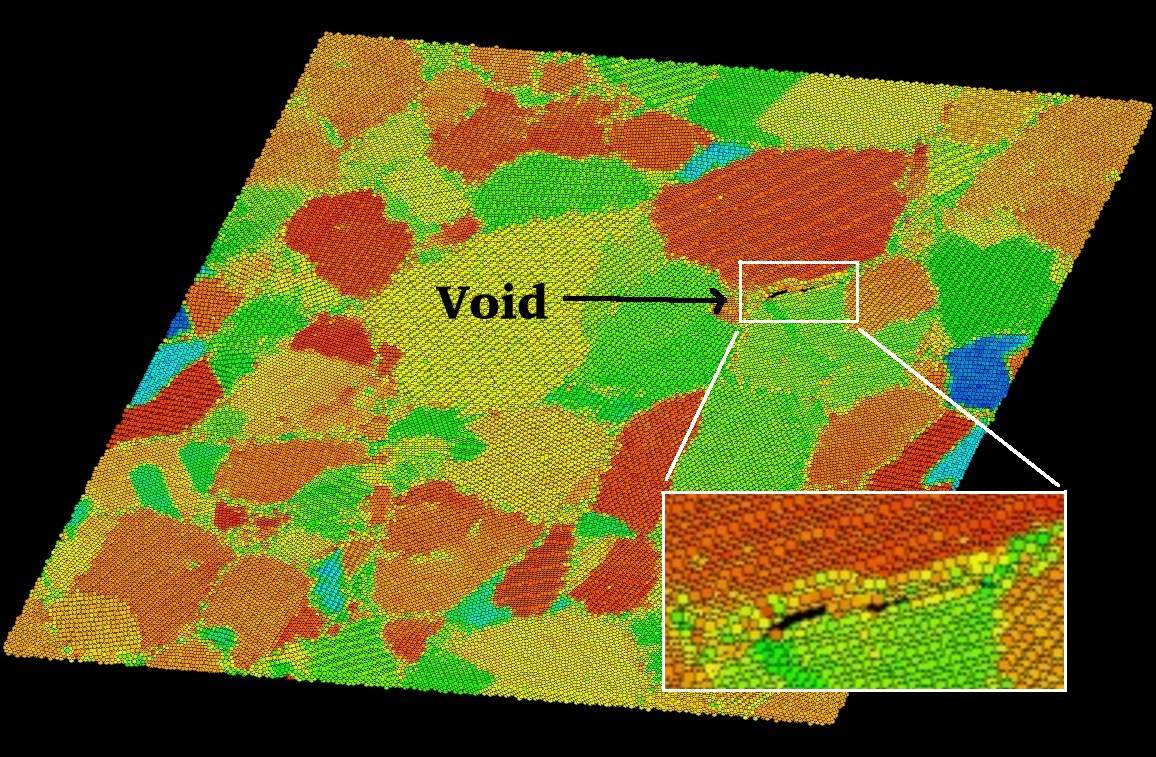}}}%
\\
\subfigure[]{
\resizebox*{70mm}{!}{\includegraphics{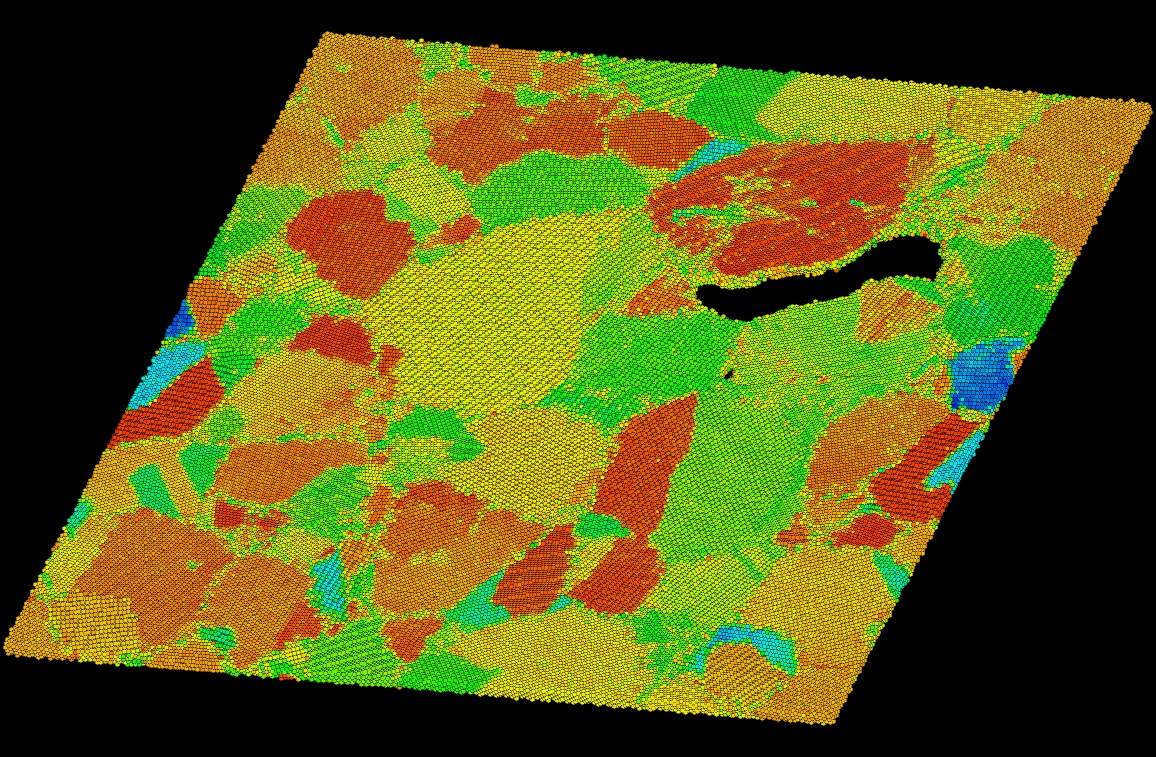}}}%
\subfigure[]{
\resizebox*{70mm}{!}{\includegraphics{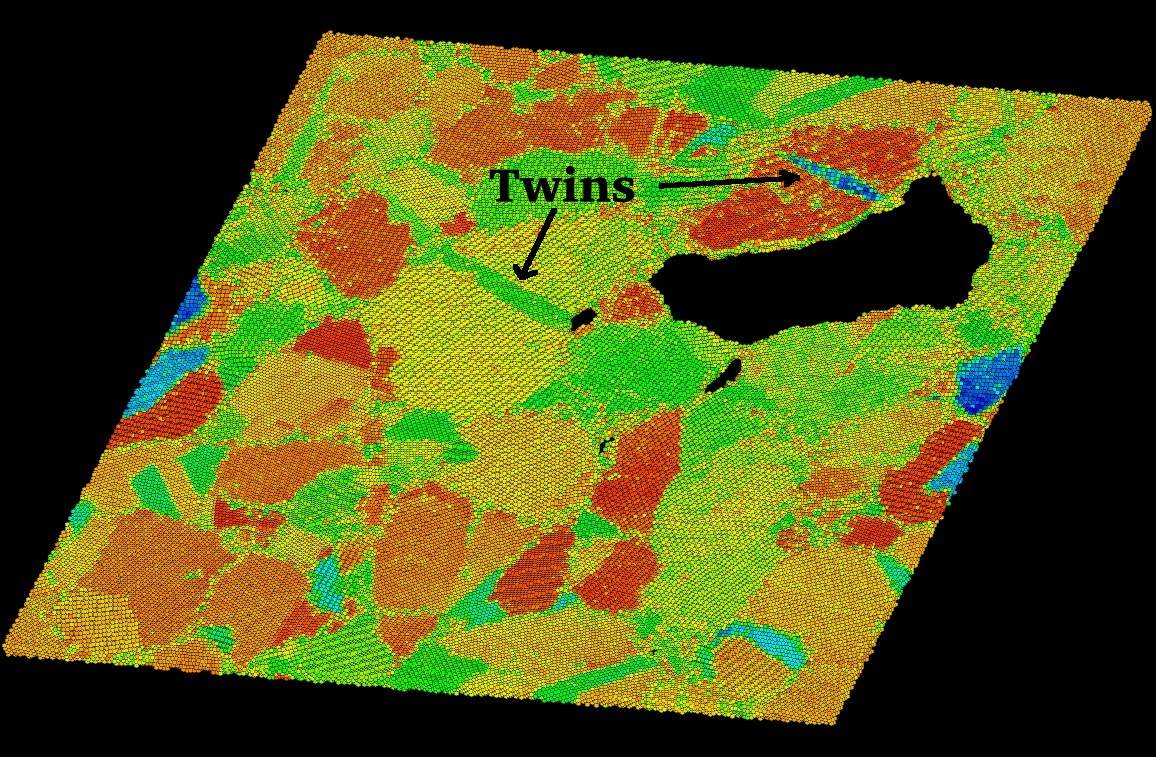}}}%
\\
\subfigure[]{
\resizebox*{70mm}{!}{\includegraphics{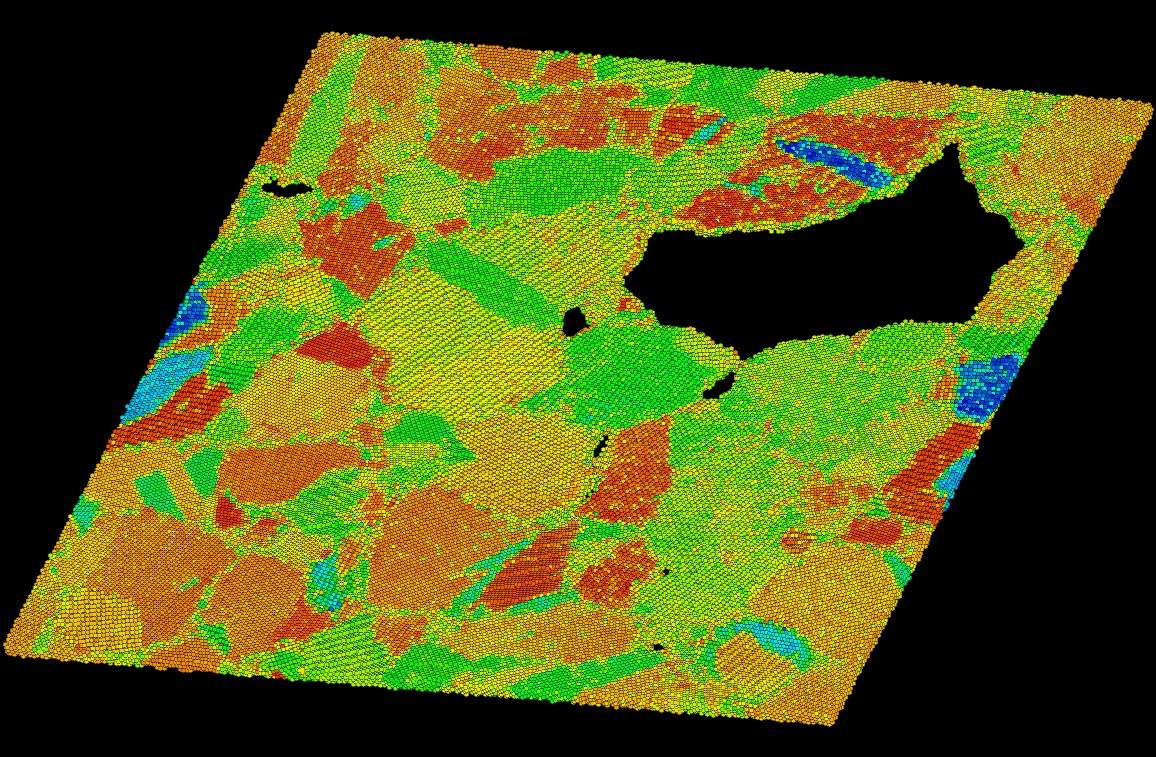}}}%
\subfigure[]{
\resizebox*{70mm}{!}{\includegraphics{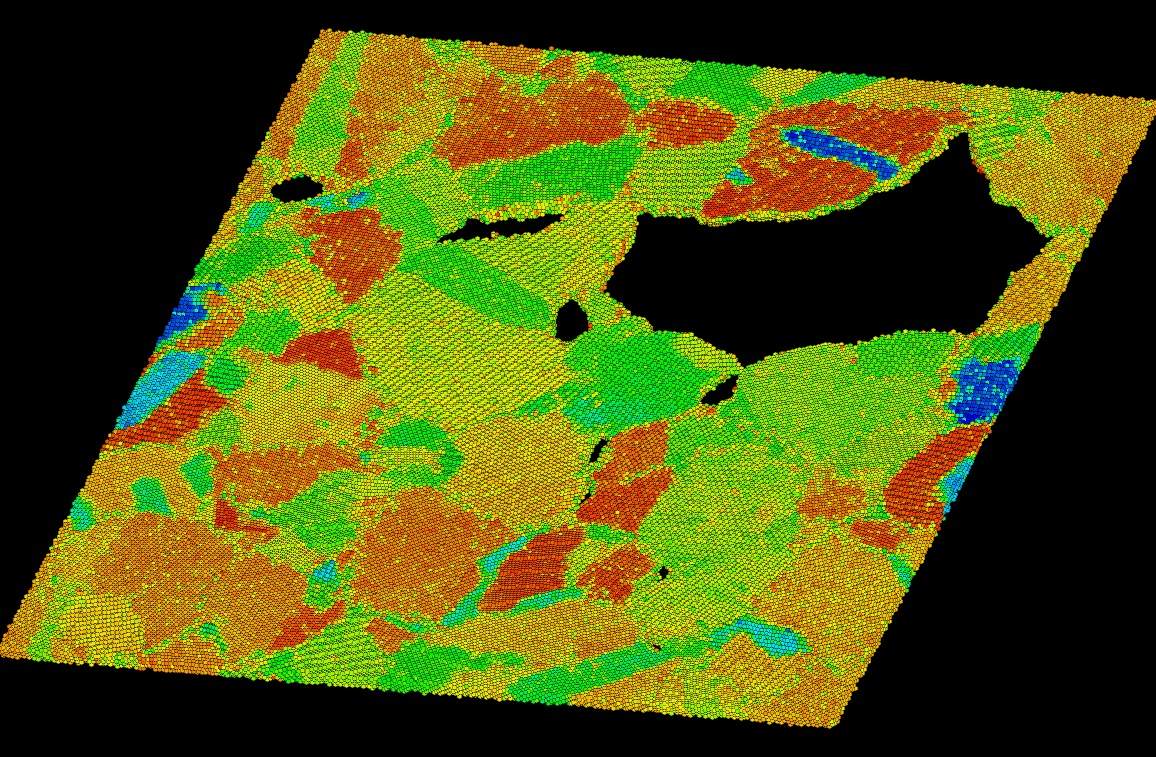}}}%
\caption{Snapshots of void nucleation and growth in an MD 
simulation of nanocrystalline Ta under tension at a strain
rate of $10^8$/s.  A 238k atom slice
through the 16 million atom cubic simulation is shown, coloured according
to the local orientation of the lattice (see text). The snapshots
correspond to different strains: 
(a) 0.0304\%,
(b) 0.0308\%,
(c) 0.0313\%,
(d) 0.0317\%,
(e) 0.0322\%, and
(f) 0.0398\%.
In panel (a) no void nucleation has taken place.
In panel (b) one void can be seen to have nucleated in
the upper right part of the image.
Through the rest of the sequence the void grows and
additional voids nucleation.
\label{fig-nanoxtal-voids}
}
\end{minipage}
\end{center}
\end{figure}

Snapshots from the $\dot{\varepsilon}=10^8$/s simulation are shown
in Fig.\ \ref{fig-nanoxtal-voids}.  The polycrystalline system
is coloured according to the quaternion for the 
local lattice orientation, as
described above.  Initially, the system is fully dense
(Fig.\ \ref{fig-nanoxtal-voids}a).  As the stress increases,
voids form at the grain boundary junctions.  A sliver of
a void is barely visible in the upper right quadrant
of Fig.\ \ref{fig-nanoxtal-voids}b.  As the box expands further,
the first void open up and additional voids nucleate
(Fig.\ \ref{fig-nanoxtal-voids}c-f).  The grains around
the large void exhibit dislocation-based plasticity
(Fig.\ \ref{fig-nanoxtal-voids}c) and then twinning
(Fig.\ \ref{fig-nanoxtal-voids}d) in order to accommodate the
void growth.

\section{Conclusion}
\label{sec-conclusion}

Simulation of the nucleation and growth of voids at the
atomistic level has provided a wealth of information about
the interaction of plasticity with the void.  The 
simulations have shown an elastic phase in which the
material stretches followed by a plastic phase in which
there is considerable void growth.  In the single
crystal simulations prismatic dislocation loops are
punched out as the void grows.  These glissile edge 
dislocation loops transport a platelet of atoms away
from the void, facilitating its growth and reducing the
elastic strain energy of the material.  The process is similar
to the dislocation processes observed in the classic
Mitchell experiments in which prismatic loops were
punched out due to the strain of a mismatched spherical
inclusion \cite{Mitchell}.
At high strain rates, the simulations
showed a propensity to twin and then fracture along
the twin boundary.  

The nature of the plasticity
associated with void growth in the bcc metals
is similar to that observed in fcc void growth
simulations previously \cite{RuddJCAMD}.  In both
cases prismatic loops have been observed to be
punched out and propagate away from the void.
At a finer level of detail, the fcc and bcc
plastic mechanisms differ substantially.
In the bcc dislocation nucleation process 
the growth occurs in bursts due to the high Peierls
barrier and some of the details of the dislocations 
are different, including a staggered nucleation process
and different loop shapes.  
There is no sign of substantial debris
formation as reported for bcc screw dislocations
in Ref.\ \cite{MarianBulatov}.

In the nanocrystalline simulations, the fully dense
metal was observed to nucleate voids at some of the grain
boundary junctions.  As these voids grew, the
grains surrounding the voids deformed plastically,
first through dislocation flow and later through
twinning.  

In these simulations we have made a number of explicit and implicit assumptions.
In the simulations reported here except for the polycrystal,
the void is the only defect in the lattice at the start of
the simulation.  In an engineering metal, a variety of defects
is present including grain boundaries, inclusions, and pre-existing 
dislocations.  Here we have focused on the case where those
other defects can be neglected, and the adhesion between 
inclusions and the matrix is sufficiently weak that the
inclusion can be modelled as a void.  The other defects can
be neglected provided their population is sufficiently
sparse, as discussed in Section \ref{sec-disl}.  
It remains for future work to go to much larger simulations
to go to lower loading rates and focus on voids nucleating
from larger inclusions surrounded by other defects and microstructural
features.

We can elaborate on the condition of only sparse pre-existing
dislocations.   The dislocation flow required in the plastic
zone around a growing void can come from the prismatic
content of dislocations emitted from the surface or 
the prismatic content of dislocations flowing in from the
surrounding material.  A far-field shear stress does not
drive prismatic flow to or from the void; it is the hydrostatic
tension with the stress concentration at the void that drives 
the prismatic flow.  So in systems that are driven hard
and in which the dislocation density is low so that the
mean dislocation spacing is much greater than the size
of the voids and void nucleation sites, it is 
dislocation flow nucleated from the void surface 
that is important for void growth.
This regime has motivated our work.

In the comparison of the peak stress in the MD simulations
with the experimental measurement of the Mo spall strength,
it was found that the MD under-predicts the stress to
initiate spall in bcc metals.  In the fcc metal copper studied
previously the agreement was much better \cite{RuddJCAMD}
The disagreement in Mo suggests that if we want
to understand the spall strength quantitatively, we either
need to simulate the full spallation process, not just 
single void growth, or we need a potential that has been
fit to tensile equation of state (EOS) data and validated, or perhaps
both.  More sophisticated potentials are available; however, the
demands of this application are challenging.  The potential
not only needs to provide an accurate description of
the EOS, elasticity and dislocation mobility in this unusual 
tensile regime, but it needs to describe the void surface
accurately, including surface structure, surface energies
and surface stresses.  All of these material properties 
affect the dislocation nucleation process. 

The question of plasticity around voids is of interest not
only when the voids grow, but also when they collapse.  
Crush down models have been developed to understand the
behaviour of voids and helium bubbles in irradiated
materials that are subjected to pressure \cite{Reisman}.
Some of the findings here may help shed light on
void collapse, as well.

These atomistic
simulations have provided a window into the dislocation
processes around the void as it nucleates and grows.  
The single void growth and nanocrystalline void nucleation
and growth processes involve complex plastic flow dynamics
governed by some fundamental processes including nucleation
at weak points such as grain boundary junctions and deformation
of the matrix accommodating void growth through a prismatic
dislocation flux.
There is still much to be learned about void growth
during dynamic fracture of bcc metals.  

\begin{center}
{\sc Acknowledgements}
\end{center}

It is a pleasure to thank Jim Belak, Eira Sepp\"{a}l\"{a}, and Laurent Dupuy
for useful discussions, as well as the earlier work on voids in fcc metals.
The initial atomic configuration for the nanocrystalline simulations was 
provided by Streitz, Glosli, and Patel \cite{StreitzSolid}.
Computer resources were provided by Livermore Computing through
a Supercomputing Grand Challenge project. 
This work was performed under the auspices of the US Department of Energy 
by Lawrence Livermore National Laboratory under Contract DE-AC52-07NA27344.

\label{lastpage}

\end{document}